\documentclass{emulateapj}

\def\mj{M$_{\rm J}\ $}
\def\rj{R$_{\rm J}\ $}
\def\etal{{et~al.\,}}

\def\teff{T$_{\rm eff}\,$}

\def\sles{\lower2pt\hbox{$\buildrel {\scriptstyle <}
   \over {\scriptstyle\sim}$}}
\def\sgreat{\lower2pt\hbox{$\buildrel {\scriptstyle >}
   \over {\scriptstyle\sim}$}}

\begin{document}

\slugcomment{Accepted to Ap.J. February 9, 2007}

\title{Possible Solutions to the Radius Anomalies of Transiting Giant Planets} 

\author{A. Burrows\altaffilmark{1}, I. Hubeny \altaffilmark{1}, J. Budaj\altaffilmark{1,2}, \& W.B. Hubbard\altaffilmark{3}} 

\altaffiltext{1}{Department of Astronomy and Steward Observatory, 
                 The University of Arizona, Tucson, AZ \ 85721;
                 burrows@zenith.as.arizona.edu, budaj@as.arizona.edu, hubeny@aegis.as.arizona.edu}
\altaffiltext{2}{Astronomical Institute, 05960 Tatranska Lomnica, Slovak Republic} 
\altaffiltext{3}{Department of Planetary Sciences and the Lunar and Planetary Laboratory, 
                 The University of Arizona, Tucson, AZ \ 85721;
                 hubbard@lpl.arizona.edu}

\begin{abstract}

We calculate the theoretical evolution of the
radii of all fourteen of the known transiting extrasolar giant planets (EGPs)
for a variety of assumptions concerning atmospheric opacity,
dense inner core masses, and possible internal power sources.  We 
incorporate the effects of stellar irradiation and customize
such effects for each EGP and star. Looking collectively at the 
family as a whole, we find that there are in fact two radius 
anomalies to be explained. Not only are the radii of a subset
of the known transiting EGPs larger than expected from previous
theory, but many of the other objects are smaller than the default theory
would allow.  We suggest that the larger EGPs can be explained
by invoking enhanced atmospheric opacities that naturally
retain internal heat. This explanation might obviate the necessity for an
extra internal power source. We explain the smaller radii
by the presence in perhaps all the known transiting EGPs
of dense cores, such as have been inferred for Saturn and Jupiter.
Importantly, we derive a rough correlation between the masses
of our ``best-fit" cores and the stellar metallicity
that seems to buttress the core-accretion model of their formation. 
Though many caveats and uncertainties remain, the resulting comprehensive 
theory that incorporates enhanced-opacity atmospheres and dense cores is 
in reasonable accord with all the current structural data
for the known transiting giant planets.

\end{abstract}


\section{Introduction}
\label{intro}

Approximately two hundred extrasolar giant planets (EGPs) have to date been discovered by
radial-velocity techniques\footnote{see J. Schneider's
Extrasolar Planet Encyclopaedia at http://www.obspm.fr/encycl/encycl.html
and the Carnegie/California compilation at http://exoplanets.eu.}.  These data yield orbital properties
and M$_p$$\sin(i)$, where M$_p$ and $i$ are the planet mass and inclination
angle, respectively. However, for the subset of fourteen EGPs that are currently known to transit
their primaries (Charbonneau et al. 2006a), the M$_p$$-$$\sin(i)$ degeneracy is broken and EGP radii (R$_p$) are measured 
as well.  With M$_p$ and R$_p$, an estimate of the (presumably) coeval stellar age, and a detailed
theoretical model that includes the effects of stellar irradiation, the general theory of
the structure, evolution, and atmospheres of irradiated close-in EGPs\footnote{We
will not use the term ``Hot Jupiters," since it is a misnomer from the point of
view of spectra and atmospheres.  The hallmarks of Jupiter's atmosphere,
cold ammonia and water clouds and methane gas, could not survive
in the extreme irradiation regime of a close-in EGP that experiences $\sim$10$^4$ times higher fluxes
and has atmospheric temperatures that are an order of magnitude hotter. In the 
atmospheres of the known transiting EGPs, carbon is in carbon monoxide, alkali metals,
not seen in Jupiter, are the predominant absorbers at optical wavelengths, and water
is steam.} can be put to the test (Burrows et al. 2000; Bodenheimer, Lin, \& 
Mardling 2001; Bodenheimer, Laughlin, \& Lin 2003; Burrows, Sudarsky, \& Hubbard 2003; 
Baraffe et al. 2003; Burrows et al. 2004; Chabrier et al. 2004; Laughlin et al. 2005; Baraffe et al. 2005).   

Recently, observers and theorists alike have focussed 
on the apparent discrepancy with published theory of 
the transit radii of some EGPs, notably HD209458b, HAT-P-1b,
and WASP-1b (Knutson et al. 2006; Bakos et al. 2006b; Charbonneau 
et al. 2006c), i.e. that these close-in EGPs  are larger
than most theories would predict.  Many would explain this anomaly
by invoking an extra heat source for the interior, perhaps caused by orbital
tidal forcing (Bodenheimer, Laughlin, \& Lin 2003), 
obliquity tides when in a Cassini state (Winn \& Holman 2005),
or penetration of gravity waves into the planetary interior that then 
dissipate at depth (Guillot \& Showman 2002; Showman \& Guillot 
2002).  Such a power source could indeed be operative, and the 
powers required are not large (\S\ref{heat}).  However, the transit radius 
of an EGP depends upon M$_p$, the stellar flux at the planet (F$_p$, \S\ref{flux}), its atmospheric
composition (\S\ref{metal_nocore}), the possible 
presence of an inner core (\S\ref{core}), its age, and the atmospheric circulation
that couples the day and the night sides (\S\ref{day}).  It also depends upon the fact
that the transit line of sight cuts the chord of the planet, and not its
radial profile (\S\ref{transit_r}).  This effect can add $\sim$3\% to $\sim$10\% to the measured
radius (Burrows, Sudarsky, \& Hubbard 2003; Burrows et al. 2004; Baraffe et al. 2003) and 
should be included in a detailed comparison with observation.

With so many determinants of a planet's radius, comparison between theory
and measurement must be multi-parametric.   
Furthermore, errors in the measured R$_p$ and M$_p$, in the ages, and in the stellar
metallicities can be large.  These introduce significant noise in the
interpretation of any one transiting EGP and are why it is more fruitful
to look broadly at the entire family.
In this way, we are able to determine the overall
systematics in the structures of close-in EGPs 
and discover trends and characteristics that would otherwise be 
obscured if we had focussed on one object at a time.  
As a result, we put less weight
on our object-by-object ``best-fits" 
than in the patterns that emerge from our
study of them collectively.  

We find that the
range of observed radii for the entire cohort of transiting EGPs
is too large to accommodate only one radius anomaly.  We
show, in fact, that some (most) transiting EGPs are {\it smaller} than past theory
would have predicted, while we confirm that some are larger than 
past theory would have predicted.  We can explain both anomalies
with 1) enhanced atmospheric opacities 
for the larger EGPs and 2) ``ice/rock" cores for the smaller EGPs. 
Such cores are predicted by the ``core-accretion" model of giant planet formation 
(Pollack et al. 1996) and ice/rock cores shrink an EGP of a given total mass
monotonically with core mass.  An extreme case is HD149026b
(Charbonneau et al. 2006b; Fortney et al. 2006).
Interestingly, we derive a rough correlation between
the inferred core masses and the parent star metallicity.  
This trend suggests their origin.

Larger atmospheric abundances, such as those measured for 
Jupiter and Saturn (\S\ref{metal_nocore}; Atreya et al. 2003; Atreya 2006; Flasar et al. 2005), 
would lead naturally to larger atmospheric opacities that retard the loss 
of heat and entropy from an EGP and delay the shrinkage of its 
radius.  However, in this paper we are not tying such enhanced atmospheric opacities
solely to enhanced atmospheric abundances/metallicities.  
This is an important point. Rather, we are suggesting that 
the atmospheres of close-in EGPs could also be altered
significantly by strong optical and UV irradiation. 
The thick hazes, absorbing clouds, and non-equilibrium 
chemical species that could thereby be produced might
lead to significant increases in the optical thickness of the 
atmospheric blanket, leading to a slowdown in the rate of loss of core heat.     
Enhanced atmospheric opacity has an effect similar to extra core power.
We note that the UV flux at the surface of the transiting planets
is as much as a factor of 10$^4$ times higher than that at the surface
of Jupiter.  Despite the much lower Jovian UV insolation, its atmosphere
contains as-yet-unidentified trace non-equilibrium species at the part in 
$\sim$10$^{10}$ level that nevertheless result in a decrease 
by almost a factor of {\it two} in its blue and green 
geometric albedos.  What might be the response in the atmosphere of 
a close-in EGP to the factor of 10$^4$ increase in UV irradiation?

Therefore, in this paper we explore the consequences for EGP radii of enhanced
atmospheric opacities.  We do this by calculating models using
solar, 3$\times$solar, and 10$\times$solar abundance atmospheres, but the latter two
should be considered ersatze for the effects of enhanced opacities 
of whatever origin.  Hence, we decouple the effects of increased 
atmospheric opacity from increased envelope heavy-element abundances.  If the
increase in atmospheric opacity were due solely to increased metallicity
and our equilibrium chemistry and opacity algorithms were correct, 
then the implied increases in the heavy-element burden of the envelope, 
if the heavy fraction in both atmosphere and envelope were the same,
could partially or wholly cancel the expansion effect of enhanced atmospheric opacity (see \S\ref{core} 
and Fig. \ref{fig:8}).  We leave open the detailed reasons for the enhanced opacities, which
could, in addition to super-solar metallicities in the atmosphere,
be non-equilibrium chemistry, errors in the default opacities,
and/or thick hazes or clouds.  In the near future, measurements of both
the reflected light and thermal emission of close-in EGPs should help to
constrain both the opacities and compositions of their atmospheres.
We note that the detection by Charbonneau et al. (2002)
of sodium in the atmosphere of HD209458b is best fit by the presence
of hazes (Fortney et al. 2003) or some additional grayish absorber.
The default theory using clear atmospheres does not explain the 
factor-of-three discrepancy (from merely solar!) in the inferred abundance of 
sodium in HD209458b's atmosphere.  

The upshot of these dual themes concerning atmospheres and cores is a theory 
that might explain all the transit radii without resorting to an 
extra power source to inflate them. Though an extra power
source is still possible, we find no simple correlation between
the magnitude of the needed power and any planetary or stellar properties. 

In \S\ref{measure}, we review the transit data and summarize their
interesting features, particularly those that demand special explanation.
Section \ref{flux} demonstrates the general dependence of transit radii 
on M$_p$ and stellar flux (F$_p$).  The latter varies by
more than an order of magnitude among the known transiting EGPs.
In \S\ref{transit_r}, we discuss the ``transit-radius" effect
that arises from the fact that we measure an impact parameter and not a radius.
In \S\ref{metal_nocore}, we present the results of our calculations
without cores for solar-opacity and 10$\times$solar-opacity atmospheres.  
These models are the baseline suite that set the stage for the discussions that 
follow \footnote{We have also calculated 3$\times$solar-opacity
models to better determine the opacity dependence of transit radii 
and provide a more comprehensive view, but do not provide the corresponding plots.}.
The higher opacity models can fit the large-radius EGPs, modulo remaining uncertainties in
their ages.  We note again that we use increased atmospheric metallicity as a convenient substitute
for enhanced opacity.  In \S\ref{core}, we discuss the effects of a central ``ice/rock"
core and calculate a range of core masses needed
to achieve better fits for the relevant EGPs.  This section motivates a possible correlation
between the inferred core masses and the stellar metallicity that might
inform models of their formation.  In \S\ref{heat}, we discuss the possible effect
on planet structure of an extra heat source and determine how much power, object by object, 
would be needed to explain the measured R$_p$s for simple models 
of planet cooling.  This section is meant merely to provide the reader
with a gauge of the range of powers that might be required should our default and
preferred set of models be shown in the future to fail 
in some crucial particular.  Curiously, we find that
inner cores are still suggested by the data even when an extra internal
heat source is present.  In \S\ref{day}, we discuss 
a major theoretical uncertainty -- the advection of heat from the
day to the night sides due to global circulation.  Atmospheric winds
at altitude and at depth remain wild cards in the general theory of EGPs. 
In \S\ref{conclusions}, we summarize our results and conclusions 
and reiterate the remaining caveats concerning the theory of EGP radii.

\section{Measurements of Close-in Giant Planets}
\label{measure}

Table 1 is a compilation of relevant data for the fourteen known transiting
planets, listed in order of increasing semi-major axis. These data include
semi-major axis ($a$), period ($P$), M$_p$, R$_p$, F$_p$, and recent
observational references.  We also provide the latest error bars for M$_p$ and
R$_p$, though when it seemed prudent we have rounded both these and the central estimates.
Note that the flux at the planet is not monotonic with orbital distance, reflecting
the fact that these EGPs orbit a variety of stars with luminosities
that span an order of magnitude.  Table 2 provides these luminosities (L$_*$),
along with other useful stellar parameters,  such as spectral type, stellar radius (R$_*$),
effective temperature (\teff), surface gravity ($g$), metallicity ([Fe/H]),
and stellar mass ($M$).  We also provide in Table 2 the system
distances; some (such as those for the OGLE set) are quite approximate.
Most of the data in Tables 1 and 2 are necessary to construct theoretical models and compare them with
the measurements.  For instance, to incorporate the effects of stellar irradiation
one needs models of the stellar spectra and luminosities.  We employed those of
Kurucz (1994), or generated our own using the atmosphere code TLUSTY (Hubeny \& Lanz 1995).
In Table 2, we include ages and their error bars, both of which should be
considered very approximate.  We list only the central guesses
of the stellar metallicities given in the literature, but ample error
bars for them should also be assumed.  The ages and the metallicities
are the least well-known quantities in Table 2 and ambiguities
in them translate into uncertainties in the interpretation of the theoretical models and 
transit data for any given object.  However, as Table 2 suggests, the 
metallicities of these EGP parents vary by a factor of $\sim$4.  The ages
probably range even more broadly.

\begin{figure}
\centerline{\includegraphics[angle=0,width=20cm]{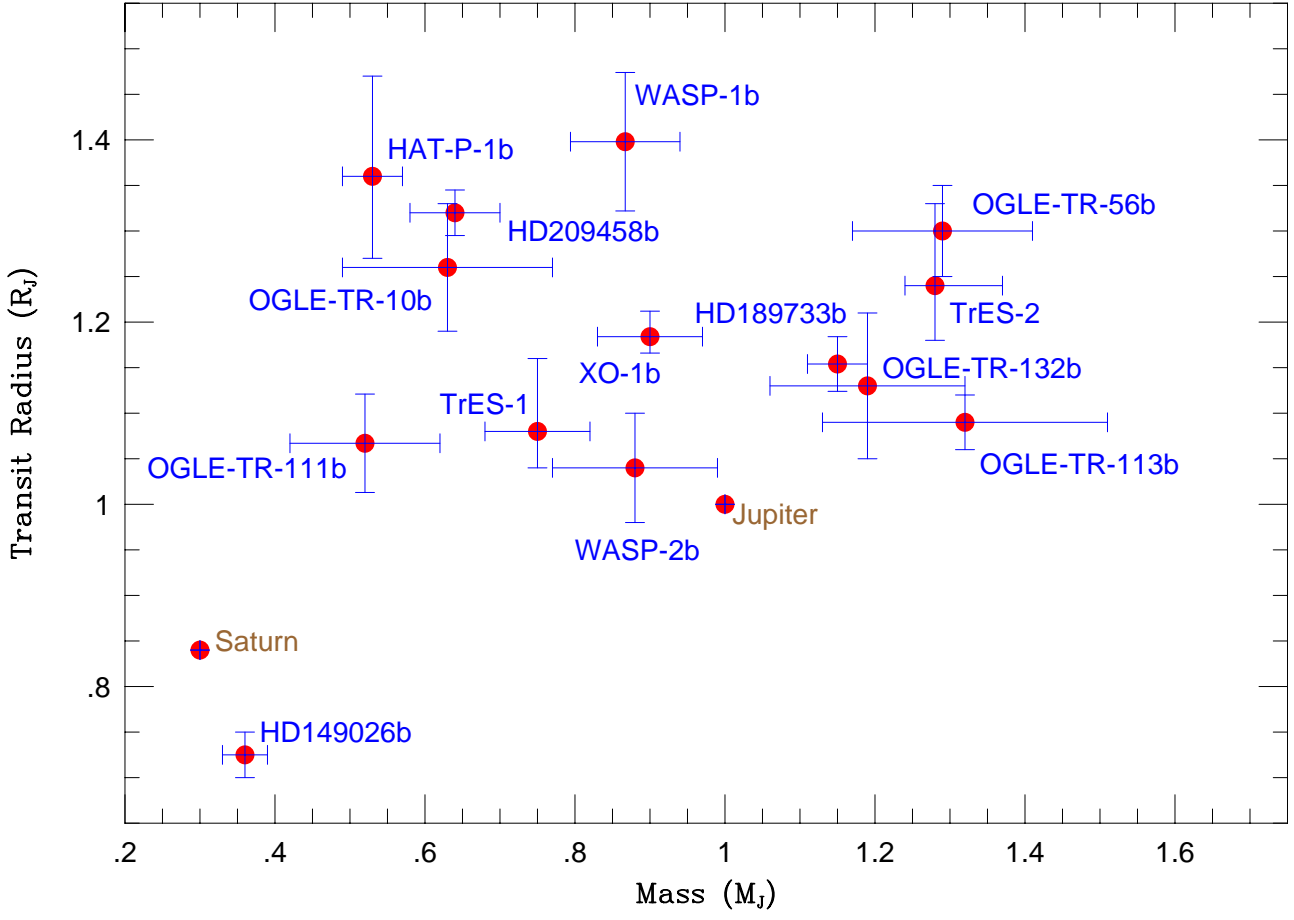}}
\caption{Transit radii (R$_p$, in \rj) of all of the irradiated EGPs listed in Table 1
versus planet mass (M$_p$, in \mj), along with published 1$-\sigma$ error bars for each quantity.  
For comparison, points for Jupiter and Saturn themselves are also shown.
\label{fig:1}}
\end{figure}

Figure \ref{fig:1} depicts R$_p$ versus M$_p$ for all the transiting EGPs given
in Table 1, along with error bars.  Jupiter and Saturn are included for context.
This figure encapsulates the basic measurements to be explained by theory
and warrants some discussion.  The first thing to note is that the spread
in transit radii is wide, $\sim$40\% for the bulk and approximately a factor
of two when HD149026b is included.  
Some have noted that there is a tendency for the larger EGPs to be orbiting 
the more massive primary stars (see Table 2).  This is most easily explained
by the fact that such stars have higher luminosities, and, hence, that their EGPs find
themselves in more intense irradiation regimes (all else being equal), but planet/star distance
and planet mass also play central roles.  In fact, the largest values of F$_p$ are 
for OGLE-TR-56b, OGLE-TR-132b, and WASP-1b, while HD209458b and TrES-2 
are in the middle of the pack (see Table 1).  As the upper envelope 
of the data in Fig. \ref{fig:1} suggests, there is a slight tendency for
the lower mass EGPs to have higher radii.  This effect is a straightforward
consequence of basic theory and is at least as important (\S\ref{flux}).

There are other apparent curiosities.  Using Fig. \ref{fig:1} and Table 1, we can compare 
subsets of EGPs with roughly the same M$_p$.  One such triplet, in order of decreasing radius,
is WASP-1b, XO-1b, and WASP-2b.  We might expect that, given this radius hierarchy,
F$_p$ would monotonically decrease from WASP-1b to WASP-2b.  However, F$_p$
for XO-1b is lower than that for WASP-2b.  HAT-P-1b, OGLE-TR-10b,
and OGLE-TR-111b constitute a similar triplet, but F$_p$ for OGLE-TR-111b is the largest
of the three, breaking what should otherwise be a monotonic trend.  Moreover,
the radii and masses of HD189733b and OGLE-TR-132b are roughly the same, yet their
F$_p$s are almost an order of magnitude different.  The most extreme case 
is HD149026b, which has the fourth highest F$_p$, but the smallest radius.
Our overall thesis is that these features can be explained, to within the error bars, not one-dimensionally,
but only after the various effects of M$_p$, F$_p$, core mass, atmospheric opacity, and age are
simultaneously addressed.

\begin{figure}
\centerline{\includegraphics[angle=0,width=20cm]{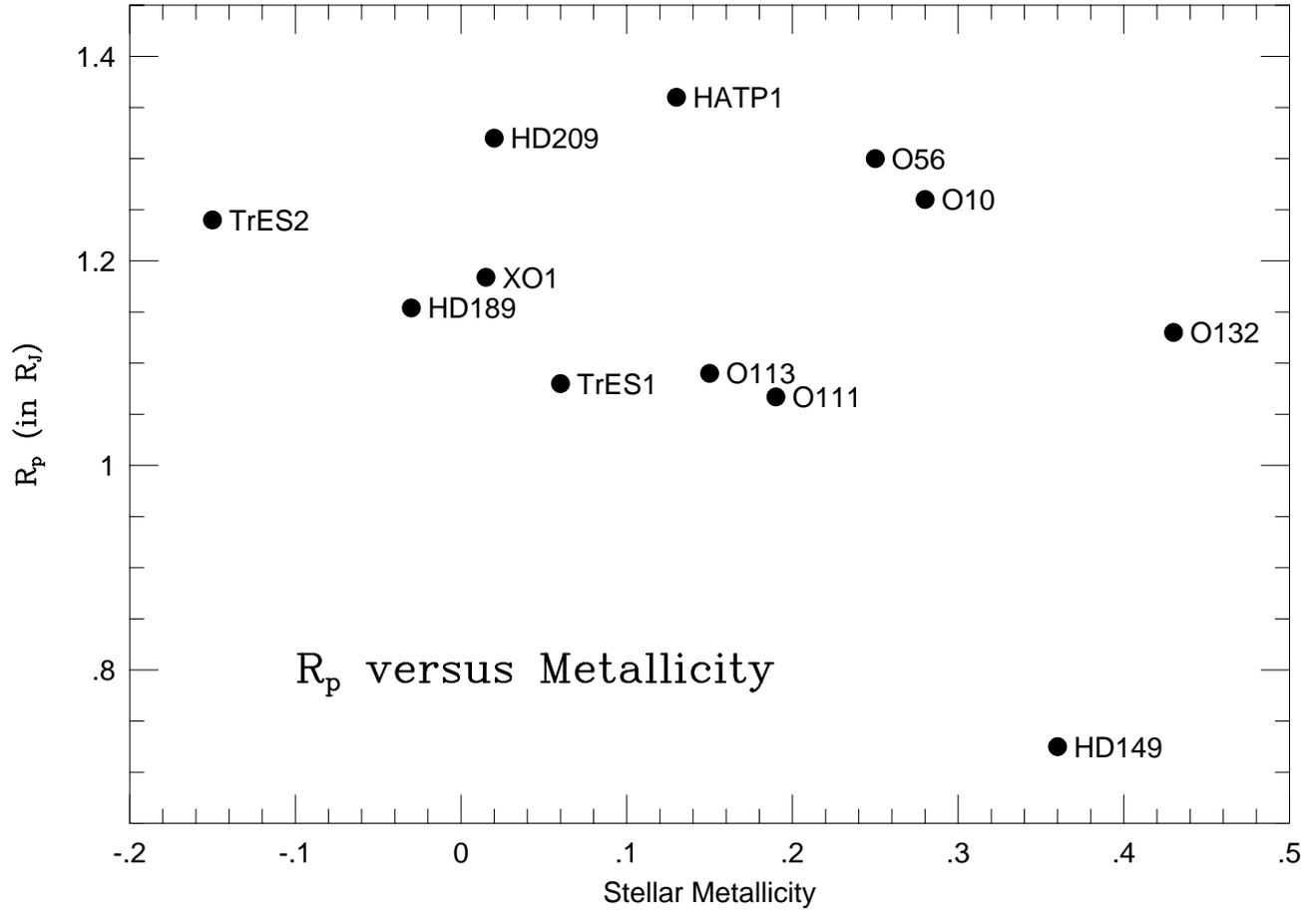}}
\caption{Measured planetary radii, R$_p$ (in \rj), versus central values
of the estimated stellar metallicities ([Fe/H]) of the transiting planets listed in Table 1,
except for WASP-1b and WASP-2b for which metallicity estimates have not yet been published.
The names for the planets are given in abbreviated form.
\label{fig:2}}
\end{figure}

Figure \ref{fig:2} depicts the dependence of the measured R$_p$ on the 
estimates of the stellar metallicity ([Fe/H], Table 2).  Error bars in
both quantities, in particular [Fe/H], will smear this plot, but the basic
relationships, if there are any, should emerge as plotted.  We see that at all
metallicities, there is a wide range of measured radii, and no clear and simple
correlation with either M$_p$ or F$_p$ (Table 1).
Curiously, there seem to be two branches (upper and lower), but 
this may be an artefact of small-number statistics.  In any case, Figs. 
\ref{fig:1} and \ref{fig:2} and Tables 1 and 2 summarize the salient 
information concerning the known transiting EGPs to be explained by theory.

\section{Dependence on Stellar Flux and M$_p$}
\label{flux}

\begin{figure}
\epsscale{1.00}
\centerline{\includegraphics[angle=0,width=20cm]{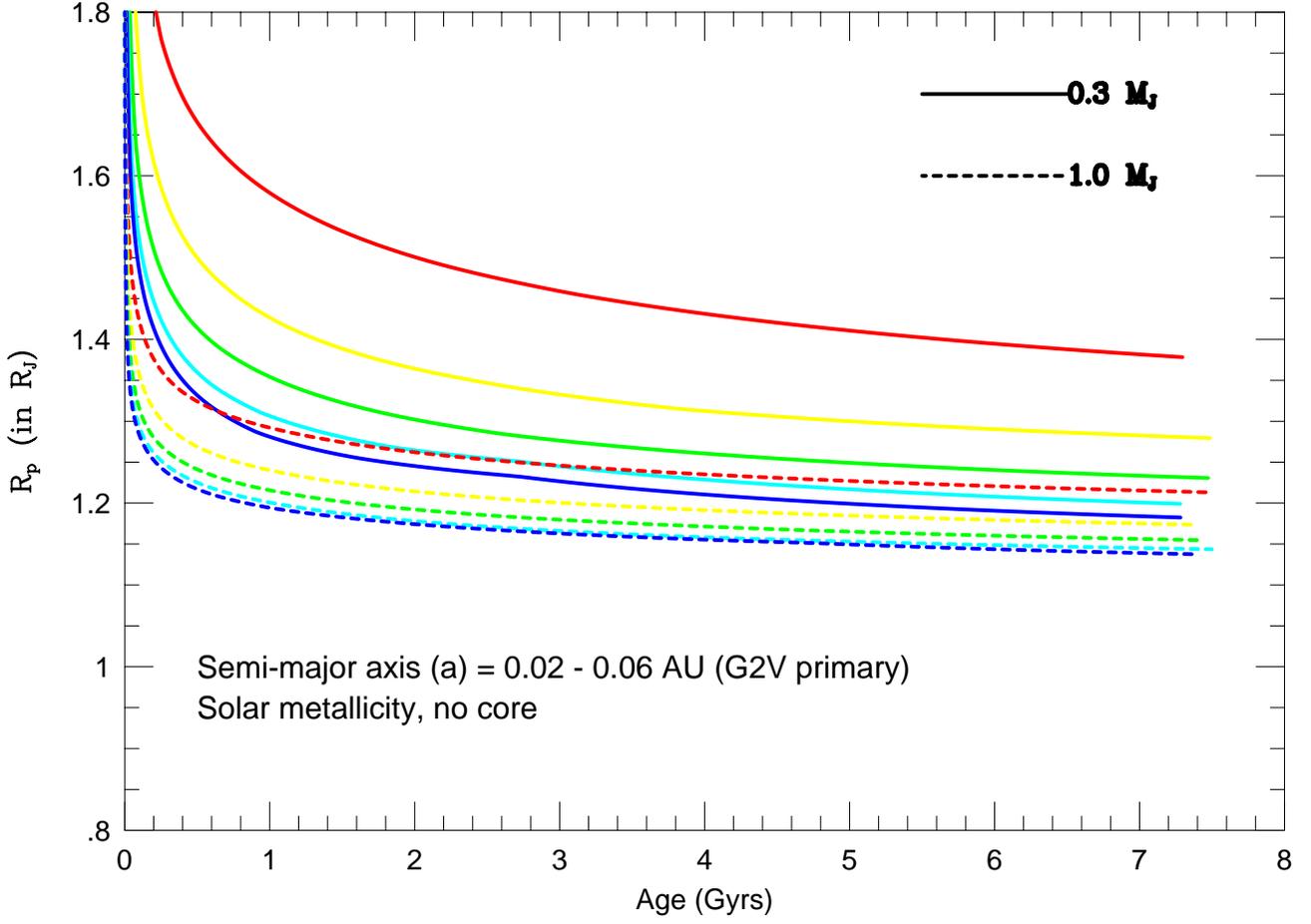}}
\caption{R$_p$ (in \rj) versus age (in Gyrs) for model planets with masses of 1 \mj (dashed)
and 0.3 \mj (solid) for different distances [0.02 AU (red), 0.03 AU (yellow),
0.04 AU (green), 0.05 AU (aqua), and 0.06 AU (blue)] from a G2V primary.
The models have no cores and assume solar metallicites when calculating the opacities.  This plot
portrays the systematic dependence of irradiated planet radii with orbital distance
for different masses.  See text in \S\ref{flux} for a discussion.
\label{fig:3}}
\end{figure}

To demonstrate the general dependence of R$_p$ upon orbital distance 
and M$_p$, we have generated Fig. \ref{fig:3}.  In it, 
we depict evolutionary trajectories for a Saturn-mass planet 
(0.3 \mj\footnote{1 \mj (Jupiter's mass) $\equiv$ 1.89914$\times$10$^{30}$ g}, solid) and a 
Jupiter-mass planet (dashed) at distances from a G2V main 
sequence star of 0.02, 0.03, 0.04, 0.05, and 0.06 AU.
These models are not per se our preferred models for any of the 
known transiting EGPs, assume solar-metallicity atmospheric abundances 
(Asplund, Grevesse, \& Sauval 2006) and opacities, do not include inner cores,
but, as do all the models we present in this paper, employ the 
well-developed boundary condition formalism of Burrows, Sudarsky, \& Hubbard (2003)
and Burrows et al. (2004).  For these, and all evolutionary calculations
in this paper, we pre-calculate grids of self-consistent irradiation boundary
conditions at 130 points that span the internal flux and surface gravity space 
(Burrows, Sudarsky, \& Hubbard 2003) likely to be traversed during the evolution of each single 
primary star/semi-major-axis combination.  During each evolutionary calculation, we 
interpolate in this grid of boundary conditions. Appropriately different
stellar spectra (see Table 2; Kurucz 1994; Hubeny \& Lanz 1995) for each 
system are employed and we set up these grids for each of the
14 known transiting EGPs and three sets of opacities (\S\ref{metal_nocore}).  
Hence, for this study we have calculated 14$\times$130$\times$3 = 5460 detailed 
spectral/atmosphere models.

We see immediately that the radius of a low-mass EGP is more sensitive to distance,
with that of a Saturn-mass EGP varying by $\sim$0.2 \rj\footnote{1 \rj 
(the radius of Jupiter) $\equiv$ 7.15$\times$10$^{9}$ cm} from 0.02 AU to 0.06 AU
and that of a more-massive Jupiter-mass EGP varying by $\sim$0.1 \rj over the same
orbital distance range.  Moreover, younger EGPs have larger radii than older representatives,
but after $\sim$1.0 Gyr all evolutionary trajectories start to flatten.  This fact
emphasizes the potential role of youth in providing large radii, and the ambiguities
that arise in the interpretation of transiting EGPs with poorly-known ages.  This is 
particularly relevant for OGLE-TR-111b, HD189733b, TrES-2, WASP-1b, and WASP-2b,
whose ages are either unknown or very poorly known.  Figure \ref{fig:3}
also shows that the timescale for radius decay is longer for lower-mass EGPs. 

\begin{figure}
\centerline{\includegraphics[angle=0,width=20cm]{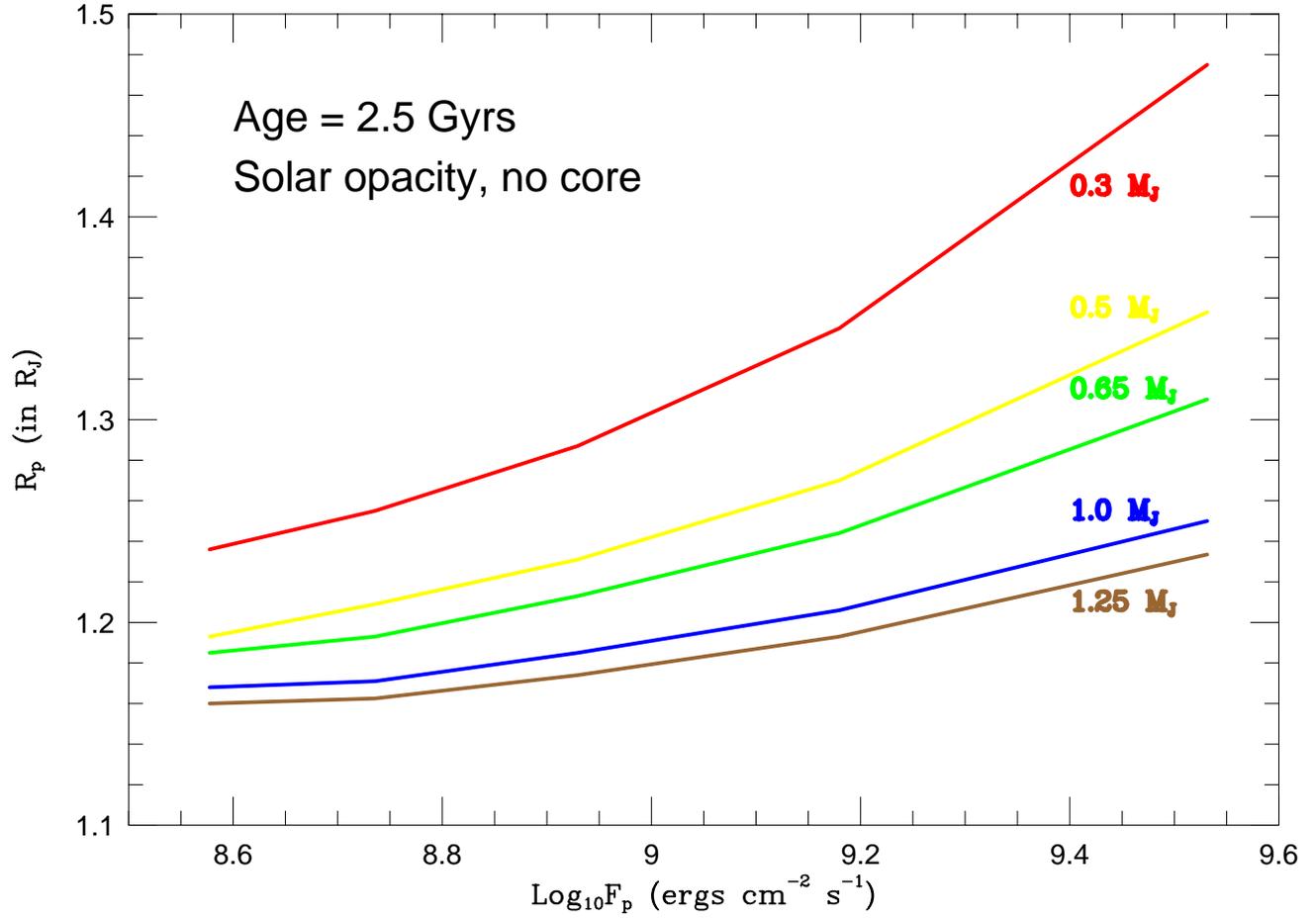}}
\caption{Solar-opacity-atmosphere/no-core model radii (R$_p$, in \rj), at an age of 2.5 Gyr, versus
the logarithm base 10 of the stellar flux at the planet (F$_p$),
in units of erg\ cm$^{-2}$\ s$^{-1}$, for a range of EGP masses from 0.3 \mj to
1.25 \mj.  This figure shows both the planet-mass and the irradiation-flux dependence
of the planet radius, at the average age of stars in the solar neighborhood ($\sim$2.5 Gyr).
See text in \S\ref{flux} for a discussion.
\label{fig:4}}
\end{figure}

Figure \ref{fig:4} continues our demonstration of the effects of 
irradiation and planet mass on R$_p$ by depicting
its direct dependence on the stellar flux (F$_p$) at the substellar 
point for the same class of theoretical models.  Roughly
one order of magnitude in F$_p$ is depicted.  Masses
of 0.3, 0.5, 0.65, 1.0, and 1.25 \mj are shown for an age of 2.5 Gyr.  This age is
roughly the mean age of stars in the solar neighborhood. Again, we see that, 
all else being equal, smaller-mass EGPs have larger radii and depend more steeply
upon F$_p$. For a 1.25-\mj EGP, R$_p$ varies for the depicted range of F$_p$s 
by $\sim$0.08 \rj, while for a 0.3-\mj EGP it varies by as much as $\sim$0.24 \rj. 
This behavior is consonant with our statement in \S\ref{measure} that the 
upper envelope of the data depicted in Fig. \ref{fig:1} has a negative slope.  
Note that the spread in R$_p$ with mass at high F$_p$
is significantly larger than at low F$_p$.  This is connected with the convergence
of the radii of cold EGPs due to the $n=1$ polytropic character of the H$_2$/He
equation of state (Burrows et al. 2001).

\section{Transit Radius Effect}
\label{transit_r}

Measuring a transit provides the impact parameter of the planet,
not its photospheric radius.  This means that the planetary limb, through
which the light from the star that defines the depth of the 
transit emerges, is at a slightly larger distance from the 
projected planet center than the canonical $\tau =2/3$ planetary radius. 
For large F$_p$s, high atmospheric metallicities, and small M$_p$,  this difference can be $\sim$5\%.
Hence, the effect should be included in any comparison with data and failure to include it will
exaggerate the apparent discrepancy with the previous theory of the radii of the biggest transiting EGPs.
For the 0.64-\mj EGP HD209458b, the effect can be larger than 0.05 \rj
(Burrows, Sudarsky, \& Hubbard 2003; Baraffe et al. 2003).

\begin{figure}
\plotone{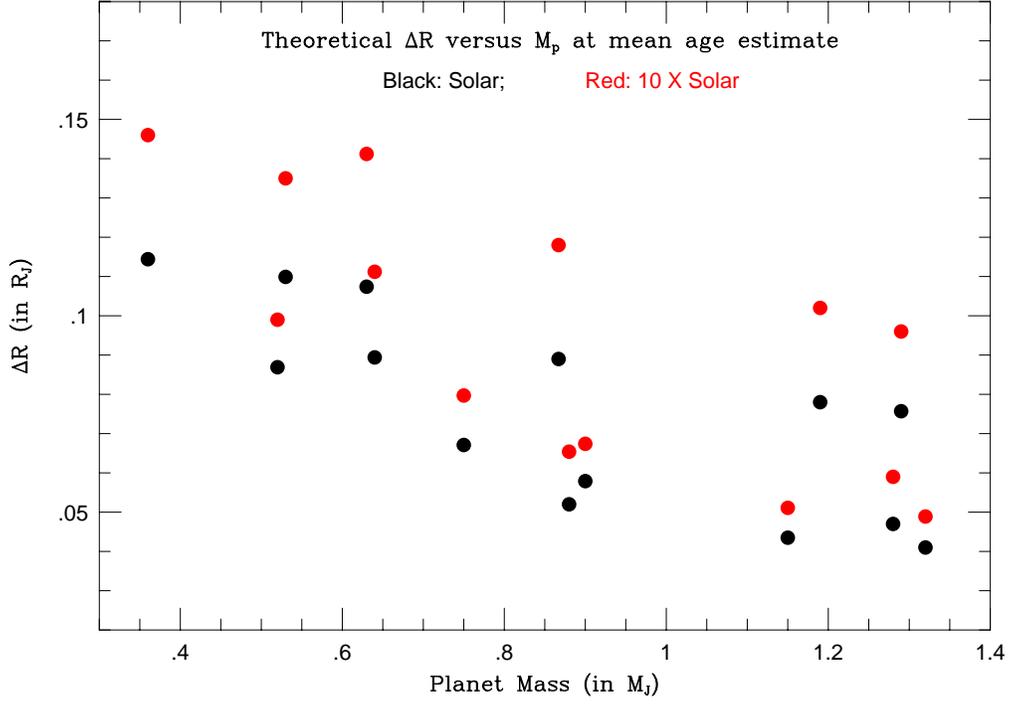}
\caption{The thickness of the radiative zone ($\Delta R$), including the transit radius effect, versus
mass for coreless models of twelve of the transiting planets listed in Table 1.  The mean molecular weight
($\mu$) used is that for pure H$_2$/He atmospheres, which is a reasonable approximation if the atmospheric heavy-element
abundance is not greatly super-solar.  Larger $\mu$s would translate into smaller $\Delta R$s.
Since age estimates for WASP-1b and WASP-2b are not published,
these objects are not included on this plot.  The central values
of the putative ages of the planets are assumed and the calculated thicknesses are
given for atmospheric opacities at solar (black) and 10$\times$solar (red) atmospheric values.
See text in \S\ref{transit_r} for a discussion.
\label{fig:5}}
\end{figure}

The wavelength-dependent optical depth, $\tau_{\rm chord}$, along a chord 
followed by the stellar beam through the planet's
upper atmosphere,  is approximately: 

\begin{equation}
\tau_{\rm chord} \sim \kappa\rho_{ph} H \sqrt{\frac{2\pi R_p}{H}}\ e^{-(\frac{\Delta R_{ch}}{H})} \, ,
\label{tau_eq}
\end{equation}
where $\kappa$ is the wavelength-dependent opacity,
$\rho_{ph}$ is the mass density at the photosphere, $\Delta R_{ch}$ is the excess radius over
and above the $\tau_{ph} = \frac{2}{3}$ radius (the radius of the 
traditional photosphere), and $H$ is the atmospheric density scale height.
The latter is given approximately by $k T/\mu g m_p$, where $\mu$ is the mean molecular weight, $g$
is the surface gravity, $T$ is some representative atmospheric temperature, and $m_p$ is the proton mass.
By definition, and assuming an exponential atmosphere, $\tau_{ph}$ = $\kappa\rho_{ph} H$ = $\frac{2}{3}$.
For $\tau_{\rm chord}$ to equal $\frac{2}{3}$, this yields

\begin{equation}
\Delta R_{ch} = H \ln\sqrt{{\frac{2\pi R_p}{H}}}\\
 \sim 5\ H  \, .
\label{tau_equ}
\end{equation}
$\Delta R_{ch}$ should be included in the theoretical radius that is compared with 
the measured transit radius.  In this paper, we include it implicitly by first calculating 
the radius of the convective-radiative boundary and then adding to it the additional 
distance to the $\tau_{\rm chord} = 2/3$ level in the corresponding detailed atmosphere model.
We refer to this additional distance as $\Delta$R (no subscript), which contains 
$\Delta R_{ch}$.  Figure \ref{fig:6} depicts $\Delta$R versus planet mass 
(Table 1) for solar (black) and 10$\times$solar (red) atmospheric opacities 
and representative coreless models of twelve of the measured transiting EGPs.  
The distance, $\Delta$R, from the radiative-convective boundary to the $\tau_{\rm chord} = 2/3$ 
level is between $\sim$0.04 \rj and $\sim$0.15 \rj for H$_2$/He-dominated atmospheres, depending mostly 
on the planet's mass ($M_p$), the stellar flux at the planet (F$_p$), and 
(weakly) its age.  As Fig. \ref{fig:5} indicates, $\Delta R$ is smaller for 
higher-mass EGPs, larger for planets experiencing higher F$_p$s (see Table 1),
and larger for higher atmospheric opacities. Concerning the 
latter, the increase in $\Delta R$ in going from solar to 10$\times$solar
ranges from $\sim$0.01 to $\sim$0.04 \rj. The contribution of $\Delta R_{ch}$ to $\Delta$R
varies from $\sim$10\% to $\sim$50\%.  Note that the numbers depicted in Fig. \ref{fig:5}
assume that the mean molecular weight ($\mu$) is not altered at high
opacity.  Even if high opacity meant high metallicity, the $\mu$ effect at 
10$\times$solar would amount to a diminution of the scale height and 
the transit-radius effect itself by no more than $\sim$20\% of the enhancement, 
and would not compensate for the corresponding increase in $\Delta$R due to the opacity effect.

\section{Models with Solar- and Enhanced-Opacity Atmospheres and No Cores}
\label{metal_nocore}

In situ and remote-sensing measurements of the atmospheric compositions of the giant
planets Jupiter and Saturn reveal that most of the dominant elements, such as
carbon, nitrogen, and sulfur, exist there in super-solar abundances (Atreya et al. 2003).  
Atreya (2006) estimates that [N/H] and [C/H] in Jupiter's atmosphere are 4-5
times solar and that [C/H] in Saturn's atmosphere is 9-10 times solar. 
Furthermore, Flasar et al. (2005) estimate that carbon in Saturn's atmosphere is $\sim$7 times solar.  Given the  
ambiguities in the interpretation of the Galileo probe results, [O/H] is problematic,
but it too is widely considered to be super-solar. Since the 
metallicities of EGP host stars are preferentially in excess of the Sun's 
(Fischer \& Valenti 2005), the idea that the atmospheres of orbiting EGPs 
are heavy-element-rich is more than just an intriguing possibility.
In addition, the excesses seen in Jupiter and Saturn are in keeping with the core-accretion model of giant
planet formation (Pollack et al. 1996), and are some of the reasons it is preferred.

It was these super-solar heavy-element abundances in the Jovian planets that first motiviated us
to explore the effects on EGP radii of enhanced atmospheric opacities.
As we suggest in \S\ref{intro}, even for solar abundances, 
strong irradiation may significantly alter the chemistry and 
opacities of the atmospheres of close-in EGPs.  Hereafter, we use 
super-solar metallicity as a substitute for enhanced opacity  
by whatever means and for whatever elemental abundance pattern and 
metallicity.  We explore the consequences for the radii of irradiated EGPs
of such opacities and compare with the corresponding results for default 
solar-metallicity atmospheres.  In the models that follow, 3$\times$solar and 10$\times$solar are to 
mean ``with heavy-element opacities that are 3 and 10 times what they would be at a given temperature 
and pressure for the canonical, unaltered solar-metallicity atmosphere."  However, note that the envelopes
of the models presented here are assumed to be pure H/He mixtures and that the effect
on the planet's radius of envelope metals is, for our purposes, ``absorbed" into 
an effect due to the core alone.  Hence, our cores ``stand in" for the core/envelope vis \`a vis their
summed effect on the planet radius (\S\ref{core}). 

Higher atmospheric opacities retain the core's heat and entropy, and this 
maintains the EGP's radius at higher values for longer times. This 
consequence of higher atmospheric gas-phase opacities (which could 
be abetted by upper atmosphere clouds; cf. Fortney et al. 2003) is 
similar in effect to that of an extra core power source
(\S\ref{heat}), but we believe that this explanation 
of large EGP radii may be more natural.  

\begin{figure}
\includegraphics[angle=0,width=9cm]{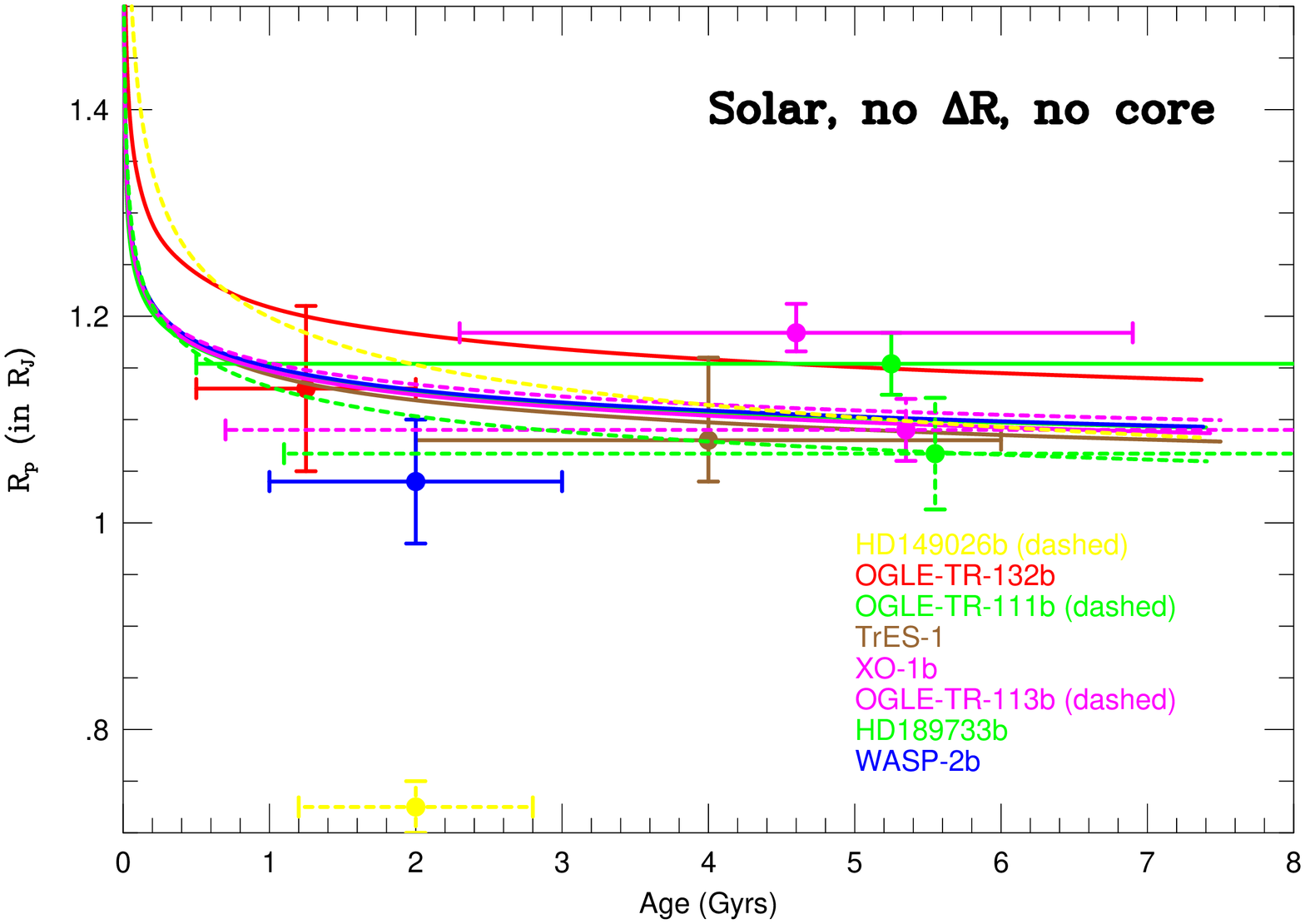}
\includegraphics[angle=0,width=9cm]{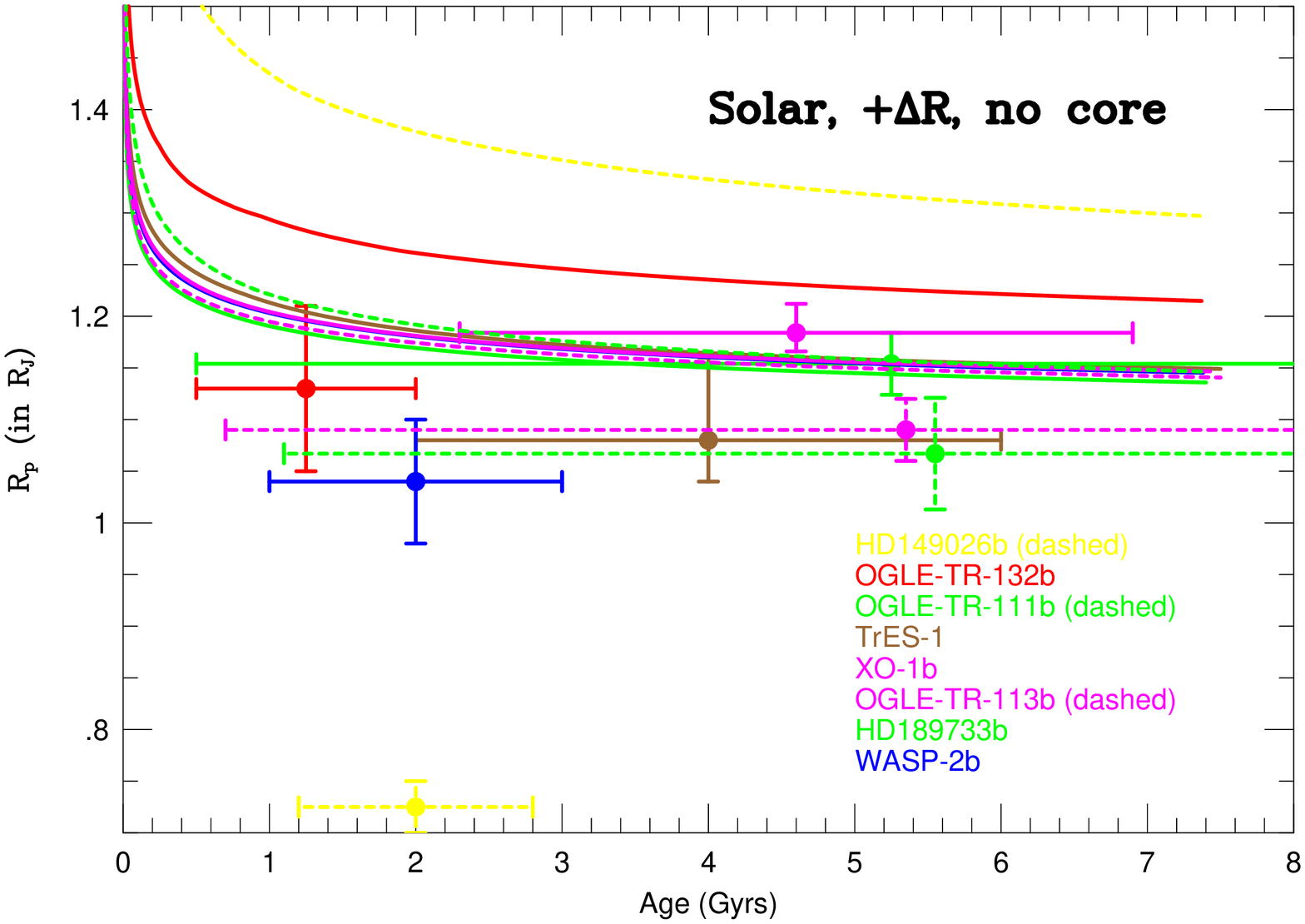}
\includegraphics[angle=0,width=9cm]{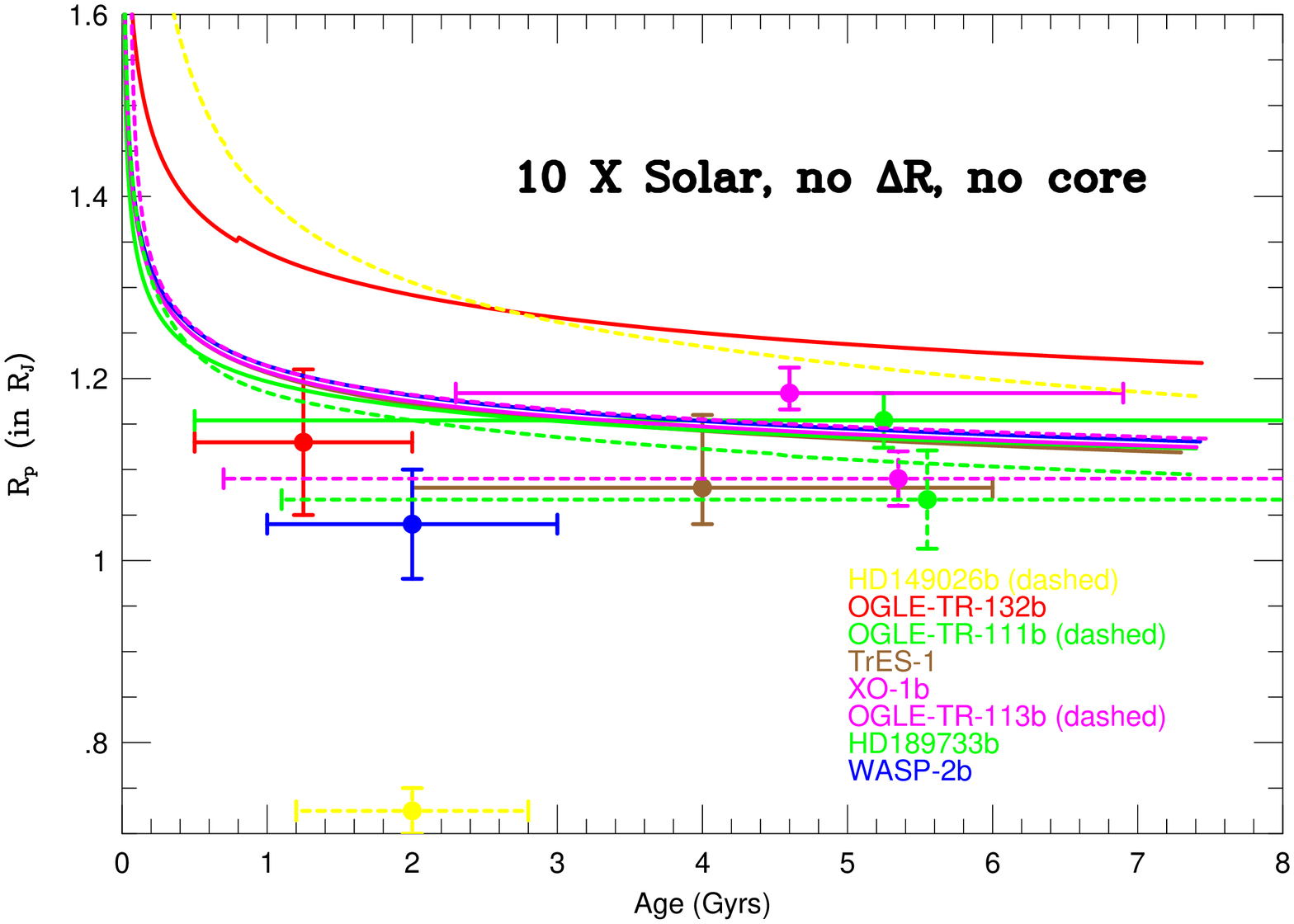}
\includegraphics[angle=0,width=9cm]{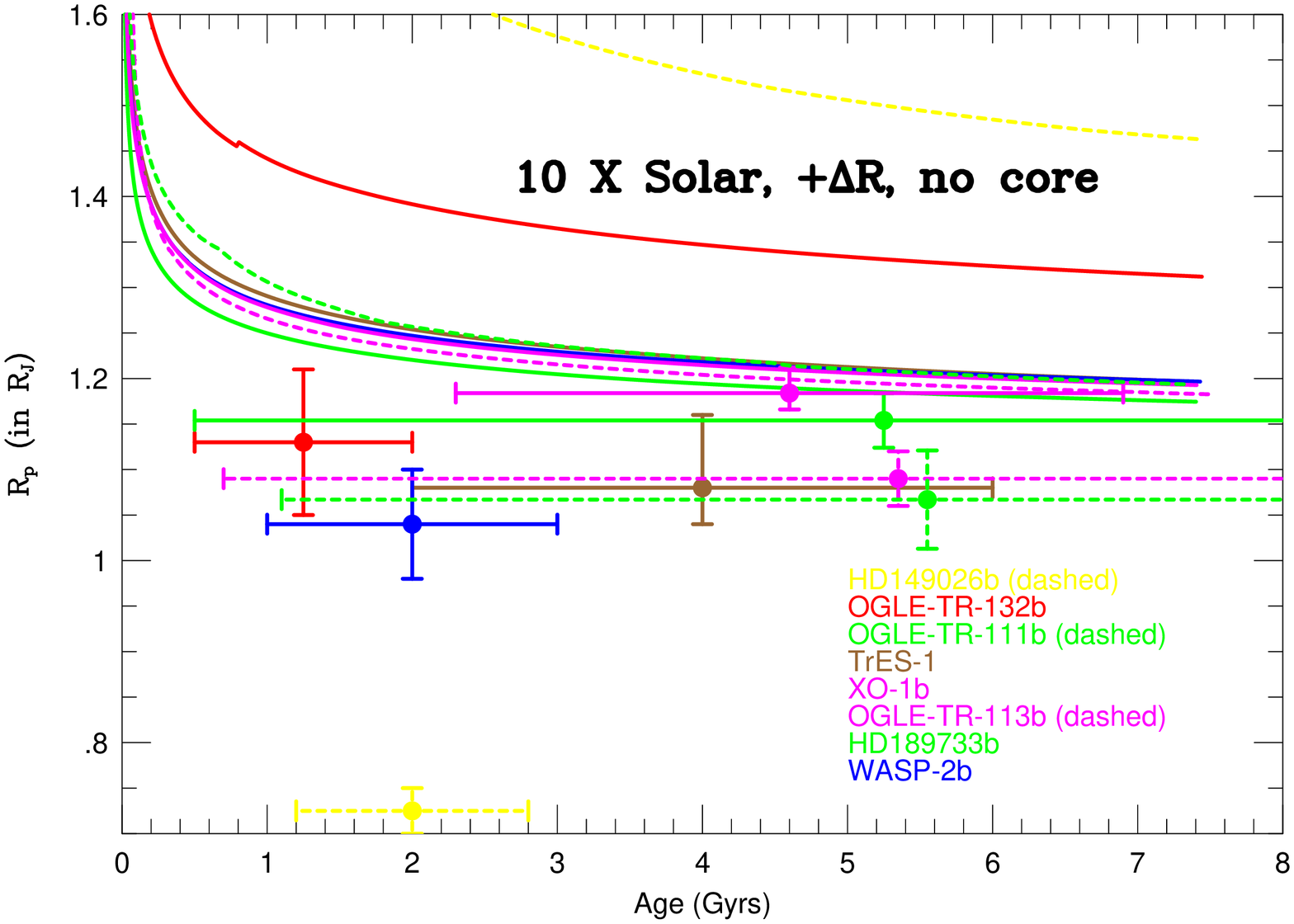}
\caption{R$_p$ (in \rj) versus age (in Gyrs) for a collection of no-core models for the
smaller transiting EGPs.  They include HD149026b (yellow-dashed), HD189744b (green), OGLE-TR-113b (purple-dashed), 
OGLE-TR-111b (green-dashed), XO-1b (purple), TrES-1 (gold), WASP-2b (blue), and OGLE-TR-132b (red).  
The top left panel is for solar opacities and does not include the
$\Delta$R term. The top right panel
is also solar, but does include the $\Delta$R term.
The bottom left panel is for 10$\times$solar opacities, but does not include the
$\Delta$R term. The bottom right panel
also assumes 10$\times$solar opacities, but does include the $\Delta$R term.
This bottom-right panel contains our default no-core/no-cloud models. 
The age of WASP-2b has been arbitrarily set at 2.0$\pm$1.0 Gyrs.
The barely-perceptible kinks near $\sim$700 Myr in the curves for OGLE-TR-132b (red)
at the lower left and right and for OGLE-TR-111b (dashed green) at the lower right are
convergence glitches in the evolutionary tracks for those models. 
See discussion in \S\ref{metal_nocore}.
\label{fig:6}}
\end{figure}

\begin{figure}
\includegraphics[angle=0,width=9cm]{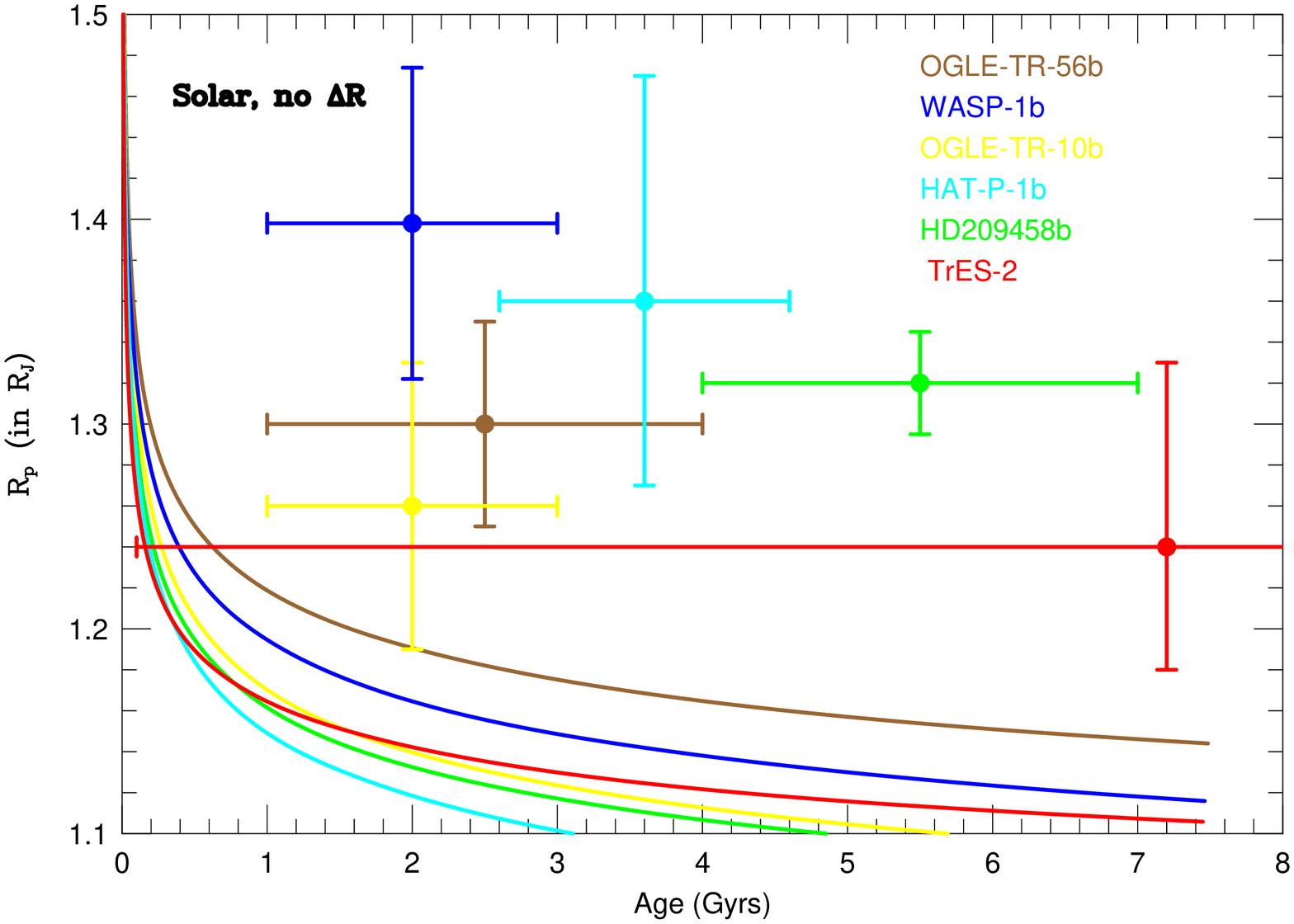}
\includegraphics[angle=0,width=9cm]{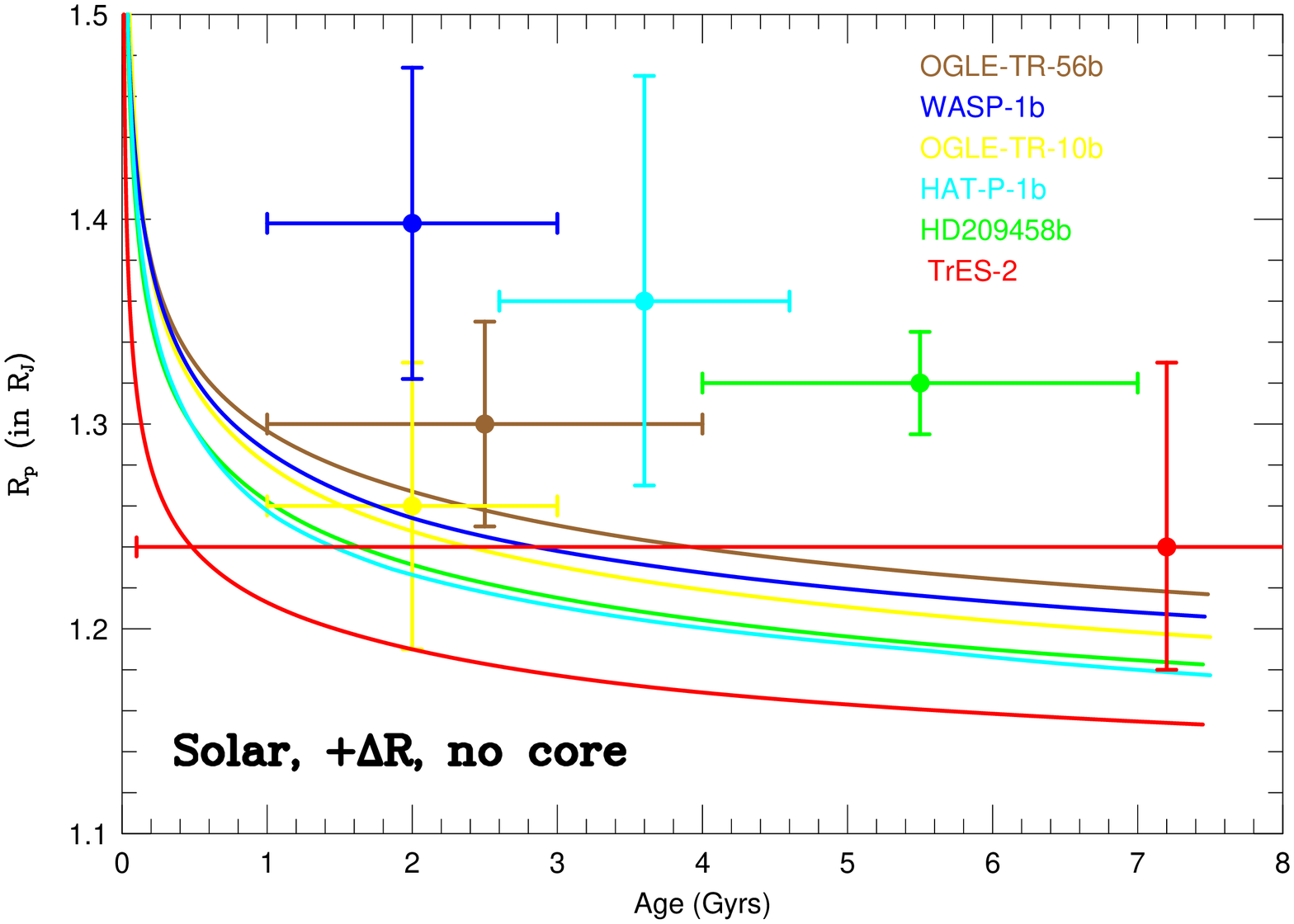}
\includegraphics[angle=0,width=9cm]{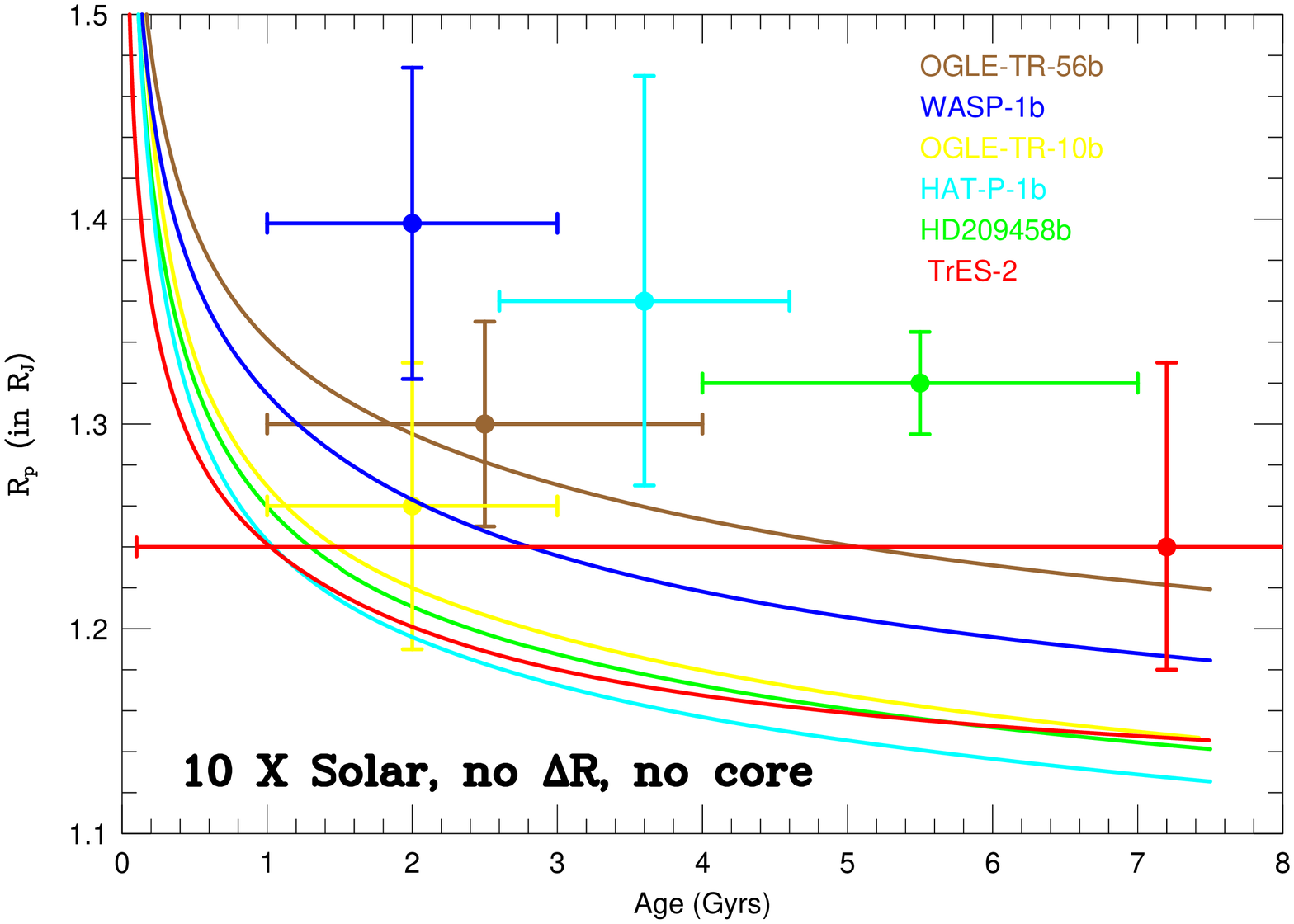}
\includegraphics[angle=0,width=9cm]{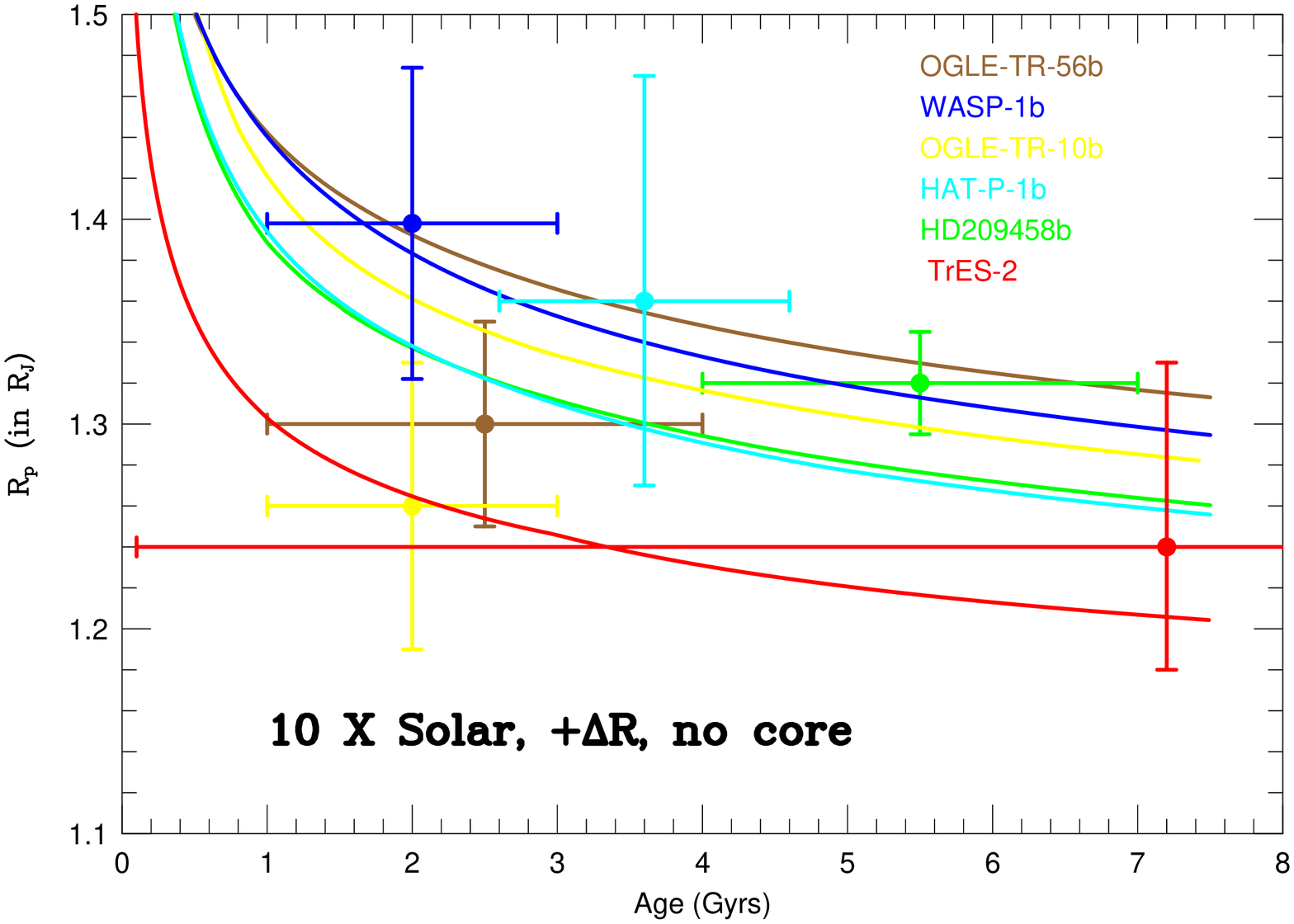}
\caption{R$_p$ (in \rj) versus age (in Gyrs) for a collection of no-core models for the
larger transiting EGPs.  They include WASP-1b (blue), HATP-1b (aqua), HD209458b (green),
TrES-2 (red), OGLE-TR-56b (gold), and OGLE-TR-10b (yellow). As in Fig. \ref{fig:6},  
the top left panel assumes solar opacities and does not include the
$\Delta$R term. The top right panel
is also solar opacities, but does include the $\Delta$R term.
The bottom left panel is for 10$\times$solar atmospheric opacities, but does not include the
$\Delta$R. The bottom right panel
also assumes 10$\times$solar opacities, but does include the $\Delta$R term.
This bottom-right panel contains our default no-core/no-cloud models.
The age of WASP-1b has been arbitrarily set at 2.0$\pm$1.0 Gyrs.
See \S\ref{metal_nocore} for a discussion.
\label{fig:7}}
\end{figure}

As stated in \S\ref{flux}, for all our calculations, we employ the evolutionary, spectral, 
atmospheric, and opacity techniques described in Burrows, Sudarsky, \&
Hubbard (2003) and Hubeny, Burrows, \& Sudarsky (2003), and 
discussed in Burrows et al. (2001)\footnote{We assume that the stellar luminosity
does not evolve with time.}. We set the redistribution factor (Burrows et al. 2004), $f$, equal to $1/4$,
and, therefore, assume complete heat redistribution at depth (see \S\ref{day}).
Figures \ref{fig:6} and \ref{fig:7} portray theoretical evolutionary trajectories
of R$_p$ versus age for coreless models of all fourteen of the known transiting
EGPs.  Model atmospheres for both solar 
opacity (top) and 10$\times$solar opacity (bottom) 
are shown and models with (right) and without (left) the $\Delta$R term are included for comparison.  
The measured transit radii and ages are superposed, along with error bars (Tables 1 and 2).
For each EGP, the color used for both model and data is the same.  Since the
ages for WASP-1b and WASP-2b are not in the literature, we arbitrarily 
set them equal to $2 \pm 1$ Gyrs.

Figure \ref{fig:6} contains eight of the smallest measured EGPs, and Fig. \ref{fig:7}
contains the other six (and, hence, the largest) EGPs. The cut between the two
sets is of no fundamental significance.  As Fig. \ref{fig:6} indicates, if we use solar
opacities, ignore $\Delta$R, and leave out a core, the left-hand-side models would fit the corresponding 
data rather well, except for HD149026b. All the coreless models of HD149026b are discrepant by
wide margins (by as much as a factor of two) and a core of substantial mass 
seems the only option (Fortney et al. 2006; \S\ref{core}).  In fact, HD149026b
is more like a super-Neptune than an EGP.  

The wide range of possible ages for some of the EGPs depicted in 
Fig. \ref{fig:6}, in particular for OGLE-TR-111b and OGLE-TR-113b,
makes interpretation a bit uncertain, particularly in the lower age range.
However, for longer ages the models are substantially age-independent.  In addition, one
can't arbitrarily ignore the $\Delta$R term and as the top right-hand panel 
of Fig. \ref{fig:6} indicates, the solar/coreless ``fits" then evaporate when including it.  Even
if the errors in R$_p$ are obliging, and data and model for a few of the eight
EGPs are reconciled, one is unlikely to be able to do this for all of them. The upshot is that even
at solar opacities coreless models for these smaller 
transiting EGPs are disfavored.  Models portrayed in the 10$\times$solar 
panels at the bottom of Fig. \ref{fig:6} are even more disfavored.  
The actual opacities of the atmospheres of these EGPs don't 
have to be as high as for the 10$\times$solar models for 
these plots to be indicative of a severe problem.  This is the first major
radius problem: many of the known transiting EGPs are too small, not
too large.  

Figure \ref{fig:7} depicts the six largest transiting EGPs in the same format as
Fig. \ref{fig:6}.  The gap with theory for solar opacities, no cores, and no 
$\Delta$R term is wide for all, except for TrES-2, if its age is quite low.  As the upper
right-hand panel indicates, including the $\Delta$R effect helps, but not enough.
However, at $\sim$10$\times$solar, the models for all these
larger EGPs start to fit rather well, the degree of fit depending centrally
upon the age and radius error bars.  In fact, for OGLE-TR-10b and OGLE-TR-56b,
their 10$\times$solar-opacity radii are on average {\it too} large.  This is true
for OGLE-TR-56b, despite its large F$_p$ (Table 1). The measured radius of HD209458b is still
a bit larger than the theory, but it is within 1.5-$\sigma$ for its central age estimate,
and better than this for younger ages.  HAT-P-1b fits well, TrES-2 fits well for a wide
range of ages. WASP-1b can fit, in particular if it is not very old (recall that its
age is unknown).  The largest opacity effects, those associated 
with an increase in radius of $\sim$0.05-0.1 \rj in going from solar to 10$\times$solar, 
obtain for the least massive EGPs (lowest M$_p$s) with the highest irradiation fluxes, 
F$_p$s (see Table 1).  For HD149026b, for which M$_p$ = 0.36 \mj and F$_p$
is the fourth highest, the magnitude of the atmospheric opacity enhancement effect is $\sim$0.2 \rj.

Therefore, we conclude that higher-opacity atmospheres
and the inclusion of the $\Delta$R term can explain 
the largest of the measured radii.  Moveover, the range of radii
among the fourteen known transiting EGPs is too wide to be explained by one factor alone.
Importantly, there is a small-radius problem as well, one that can not be solved by
an extra heat source.  We next show in \S\ref{core} that ice-rock cores for 
almost all the known EGPs, with smaller cores for the largest EGPs, are 
indicated.

\section{Effect of a Central Core}
\label{core}

In the core-accretion model of giant planet formation (Pollack et al. 1996), a
mass of ice and rock accummulates until it achieves a critical mass.
This critical mass then nucleates rapid gas accretion and the giant
planet grows to its final mass at the expense of the surrounding 
protostellar nebula. Such a two-step process is suggested because 
nebular temperatures are estimated to be too high for the inferred disk areal mass densities
to allow direct gravitational instability by the Toomre condition (Boss 1997, 2001), 
akin to the Jeans criterion for star formation.  Importantly, there
is direct evidence for the presence of a $\sim$15 
Earth-mass\footnote{One Earth mass is equal to 5.98$\times$10$^{27}$ g.} core in Saturn
and some evidence for a similar core in Jupiter (Guillot \& Saumon 2004).   The ice giants 
Neptune (17.1 Earth masses) and Uranus (14.5 Earth masses) are thought 
to be such nuclei that may have been starved of gas at birth by the
low-density neighborhood in which they were born.  

In all cases, for a given total planet mass, the presence of a core 
shrinks the total radius of an EGP. 
We numerically incorporate such cores into our models by placing a compressible ball of
olivine in the center of the model planet.  For each model, the core mass (M$_c$, in Earth masses,
as per convention) is set and pressure continuity between the solid core and the gaseous envelope
is ensured throughout the evolution.  The ANEOS equation of state (Thompson \& Lauson 1972) is
used for olivine and the Saumon, Chabrier, \& Van Horn (1995, SCVH) equation of state is used for
the H$_2$/He envelope.  In these calculations, we assume that the specific heat 
capacity per mole of the solid cores is the same as derived using the SCVH equation of state.
What the actual specific heats and entropies of the core and heavy-element component
of the envelope are is an important open issue.  If the core has a high thermal inertia,
this can delay the cooling of the planet and the shrinkage of its radius.  Conversely,
if the heat capacity of the core is smaller than that of H/He mixtures, 
large core models will cool down slightly more quickly
than our corresponding models, resulting in slightly smaller planet radii.
The zero-pressure density of olivine is $\sim$3.2 g cm$^{-3}$, significantly higher
than the average density of EGPs (Charbonneau et al. 2006a).  This is the point.
If we replace the olivine with ices or ice/rock mixtures 
the results vary slightly, but not qualitatively.  Reliable equations of state   
for heavy-element-rich gaseous envelopes that could constitute
most of the planet's mass are still not available, so we assume
that these envelopes are dominated by H$_2$/He mixtures. We have set
the helium mass fraction equal to 0.25.  Some think that whether the heavy elements
are in the core or the envelope, their effect on R$_p$ is the same.  This has not
been shown, but one can consider the inner core masses with which we deal as substitutes for
the total heavy-element burden in the planet.  It is the systematics in the group
of known transiting planets for which we are looking and the favored parameters           
of each EGP are bound to improve significantly with time.

Figure \ref{fig:8} plots theoretical total
radii as a function of core mass, M$_c$, for the estimated ages of 
OGLE-TR-10b, OGLE-TR-56b, HD189733b, and XO-1b (Table 2).
These are merely representative.  The lines in Fig. \ref{fig:8} 
are for solar, 3$\times$solar, and 10$\times$solar
models.  The measured radii of these transiting EGPs are given as dots and the
1-$\sigma$ radius error bars are indicated with vertical lines. 
The dots are placed arbitrarily along the horizontal direction 
at core masses equal to the mass fraction represented by 3$\times$solar metallicity times the   
total EGP mass and the rightmost extent of the horizontal ``error bars"
is placed at the corresponding 3$\times${\it stellar} metallicity masses.  If
the central value of the estimated stellar metallicity is below
solar (e.g., HD189733b), the horizontal line is truncated at the dot.
As Fig. \ref{fig:8} indicates, a core mass of $\sim$20 Earth masses
can shrink an EGP by $\sim$0.05$-$0.1 \rj.  EGPs with smaller total masses (such as for 
OGLE-TR-10b and HD149026b) manifest a steeper drop in radius with increasing M$_c$.
Also, while only small cores are indicated for XO-1b and HD189733b,
for OGLE-TR-10b the core mass derived for an atmospheric 
metallicity of $\sim$3$\times$solar (if metallicity and opacity 
were tied) is also the preferred core mass, i.e., the center 
of the yellow cross intersects the dashed line.  This is not often the case
in our current model set, since we have decoupled envelope metallicity from
atmospheric opacity.  However, note that, as Fig. \ref{fig:8} shows, since the intercept
of the solar-metallicity lines with the y(radius)-axis is often (though not always) below the radius positions for,
e.g., the 3$\times$solar metallicity dots for the dashed 3$\times$solar lines, 
the increase in the radius due to increasing the metallicity in the atmosphere
is often slightly larger than the decrease in the radius due to a possible corresponding increase in the 
metallicity in the envelope.  Moreover, adding ices to constitute a true 
``ice/rock" core, substituting for the pure olivine core we now assume,
should slightly favor larger radii, but we leave this to future studies.
So, higher metallicities overall can still
be an important part of the solution to the large-radius problem.  Nevertheless, more
work on the envelope equation of state for arbitrary heavy-element fractions
is still clearly needed.

Table 3 lists in bold the approximate core masses 
that provide model fits with solar, 3$\times$solar,
and 10$\times$solar atmospheres for 
each of the fourteen EGPs.  We have rounded the best-fit 
core masses to the nearest convenient number.  In 
parentheses in each column, to the left and the right
of the bolded values, are the best-fit core masses for radii that are $\pm$1-$\sigma$ 
from the central radius estimate (Table 1).  Hence, the right value is for the larger
(+1-$\sigma$) radius and the left value is for the smaller (-1-$\sigma$) radius.    
If there is no good fit, or M$_c$ would have had to be negative, we print ($\cdots$).  
``$0$" means close to zero, but could be a bit larger.  Since we have no 
age estimates for WASP-1b and WASP-2b, we leave the corresponding 
rows for them empty in anticipation of future data.  
For HD149026b, we provide only central estimates 
and for HD209458b (for which we provide no estimates) 
the best fits require that the actual transit radius be 
beyond its 1-$\sigma$ radius error bars. All these core masses are 
derived at the central age estimates given in Table 2. Since 
in most cases these ages are quite uncertain, the actual ages could yield 
very different core mass estimates. For instance, if 
an EGP's age is significantly younger, the predicted radius without a 
core would be higher (see Figs. \ref{fig:6} and \ref{fig:7}). In that case, compensating 
for the resulting larger radius deficit would require a larger core mass, 
all else being equal. 

We see in Table 3 that larger core masses are required in models with higher atmospheric
opacities, with a swing of $\sim$20-30 Earth masses from solar to 10$\times$solar.
We also see that the range of theoretical values for M$_c$ is very large, from zero
to $\sim$100 Earth masses.  Furthermore, Table 3 suggests that the canonical 
``15-Earth masses" that works for our solar system giants might be disfavored as the generic giant planet core mass.  
Moreover, we note that the high-M$_p$ OGLE-TR-132b and the low-M$_p$ HD149026b
both require very large cores, though superficially the radius of OGLE-TR-132b
might not have seemed anomalous.  For HD149026b, with a small M$_p$, a large F$_p$, and a small R$_p$,
the conclusion that a large core is required is unexceptional.  
But for the more massive OGLE-TR-132b, with the highest
F$_p$ of the family, it is intriguing that a very large M$_c$ of comparable magnitude 
may be required.  We draw a similar conclusion for OGLE-TR-113b, which is the most 
massive of the set and has a modest F$_p$, but may require a core mass of 60-80 Earth masses. 

\begin{figure}
\includegraphics[angle=0,width=15cm]{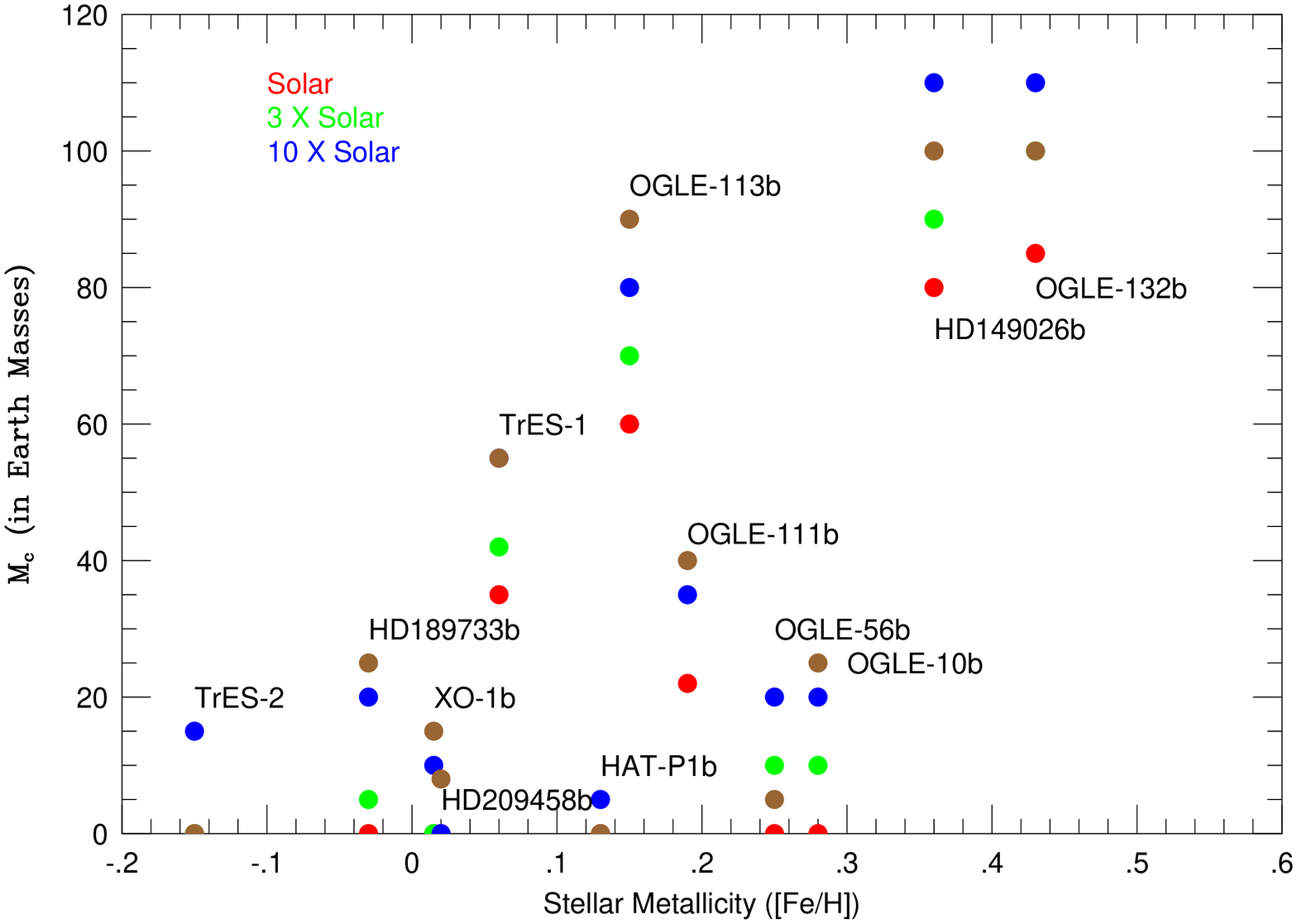}
\caption{Estimated core masses, M$_c$,  (dots, in Earth masses) versus 
the measured stellar metallicities for twelve of the transiting EGPs listed in
Table 1. The values are taken from Table 3, where the mean estimated
ages of the systems (Table 2) are assumed.  For each EGP, values for solar (red), 
3$\times$solar (green), and 10$\times$solar (blue)
opacities are given.  Since stellar metallicities for WASP-1b and WASP-2b are
not published, these EGPs are not included on this plot.  Note that despite the clustering
at low core masses for the large-radius exemplars (particularly HD209458b 
and HAT-P1b) and the low core masses for the moderate-stellar-metallicity 
EGPs OGLE-TR-56b and OGLE-TR-10b, there is a 
roughly linear (or, better, monotonic) correlation between 
metallicity and estimated core mass.  The gold points indicate the approximate
core masses necessary to fit the measured radii when the EGPs in question boast an extra internal 
power equal to a fixed 0.3\% of the corresponding L$_p$ (Table 4).
For these last models, solar-opacity atmospheres, but no irradiation or $\Delta$R terms, 
are presumed, i.e., these are toy isolated models with cores and internal heat sources.  
See text in \S\ref{core} and \S\ref{heat} for discussions.
\label{fig:9}}
\end{figure}

What patterns emerge from this theoretical study and Table 3?  Figure \ref{fig:9} 
plots the parent stellar metallicity versus the theoretical core masses given in Table 3 for  
twelve of the known transiting EGPs. The different dots for each planet 
are for the three different atmospheric opacities. In this plot and 
in Table 3, the dependence of M$_c$ on atmospheric opacity for each of the EGPs is seen 
to be less important than the wide spread in M$_c$ from object to object.
The most important feature to emerge from Fig. \ref{fig:9} is that M$_c$ seems
to increase with [Fe/H].  Based upon their preliminary analysis, Guillot 
et al. (2006) suggest a similar correlation.  Those stars with the 
lowest [Fe/H], such as HD189733, XO-1, HD209458, and the parent of TrES-2, all 
seem to be orbited by EGPs that require small cores.  Those 
stars with the largest values of [Fe/H], such as OGLE-TR-132 and HD149026,
seem to house EGPs that require the largest cores.  A ``straight" line can be drawn
through the points, suggesting a correlation between inferred core mass
and stellar metallicity.  Moreover, around solar values of the stellar metallicity,
the suggested core masses are in the solar-system regime, paralleling 
Jupiter and Saturn.  Finally, on Fig. \ref{fig:9} at low stellar metallicity no large
cores are derived, and at high stellar metallicity no small cores are derived.
To be sure, there are deviations from this simple picture,
such as OGLE-TR-56b and OGLE-TR-10b, but these points are derived using central
values of the poorly known ages and stellar metallicities.  If
the metallicities and/or ages of OGLE-TR-56 and OGLE-TR-10 are slightly lower,
the corresponding points will move up and to the left, into the trend line.
Similarly, if we derive their core masses using the upper 1-$\sigma$ radii, the corresponding
dots will shift upward on Fig. \ref{fig:9}.  However, it is also not altogether unreasonable to
expect some scatter in giant planet formation and in M$_c$.

On Fig. \ref{fig:9}, for each single object one point separately is not very suggestive, but plotted
together they collectively indicate a correlation that hints at their 
origin. At the very least, super-solar and super-stellar heavy-element 
abundances in the interiors of these planets, if not the presence of cores per se, 
are strongly suggested. Hence, we offer Fig. \ref{fig:9} as tantalizing evidence for the 
presence of dense cores and/or heavy-element-rich envelopes in 
EGPs and, therefore, for the core-accretion model of giant planet formation.

\section{Effects of Extra Heat Source in Interior}
\label{heat}

Many workers have sought to explain the large radii of transiting planets such as HAT-P1b, WASP-1b,
and HD209458b by invoking an extra power source in the planet's interior (Bodenheimer, Laughlin, \& Lin 2003;
Guillot \& Showman 2002; Winn \& Holman 2005; Chabrier et al. 2004; Charbonneau et al. 2006c). Such
a power source would maintain the entropy in the core, and hence its radius, by 
compensating in part for radiative cooling at its periphery from all quadrants.  
As our discussions in \S\ref{metal_nocore} and \S\ref{core} indicate, we don't 
prefer this solution, and in fact conclude that there are two radius problems, only one
of which could be resolved with an extra core heat source.  Nevertheless, it is useful to estimate
the magnitude of the power required for each transiting EGP to affect its measured radius.
The goal is to determine whether the requisite power could be correlated
with some other system parameter, such as intercepted stellar power, L$_p$.  In Table 4,
we provide such estimates for the transiting EGPs for two cooling models. The first
(labeled ``Power (Iso)") ignores stellar irradiation completely and assumes
the object can otherwise be considered isolated (see also Chabrier et al. 
2004).  The central value of the measured radius (Table 1) is assumed to be the 
target of the fit and $\Delta$R is not added.  Solely 
for the purposes of illustration, the atmospheres have solar opacities.  
We see in Table 4 that between 0.45\% and 0.005\% of each EGP's L$_p$ 
would be called for.  The characteristic variation is 
a factor of ten.  This needed variation from object to object makes unclear the origin
of such a power source.   

The second model (``Power (Solar)") also assumes that the atmospheres 
have solar composition and drops the $\Delta$R, but includes the effect of 
stellar irradiation with our default redistribution parameter (\S\ref{day}).  These models are
the solar-atmosphere/no-$\Delta$R models described in \S\ref{metal_nocore}, but with an extra power
source.  In this case, the range of fractions of L$_p$ is more narrow,
between 0.01\% and 0.05\%, and a factor of ten smaller than for the ``Power(Iso)" 
model set, reflecting the effect of irradiation.  Note that for more than half 
the models in this model set an extra heat source would make the radius fit worse, not better.
Other atmospheric opacities/metallicities could have been used in this 
illustrative study, but the qualitative results would have been similar.
To further demonstrate the dependence on core power of the evolution of R$_p$, 
Figure \ref{fig:10} depicts such trajectories for two representive 
EGPs, HD209458b and HAT-P1b, for both ``Power (Iso)" and ``Power (Solar)" assumptions
and for a variety of core powers.   

While it is noteworthy that the fraction of L$_p$ needed to modify
R$_p$ in a measureable way is quite small, no natural mechanism 
and no systematic reason for significant variation from object to object suggest themselves.  
Nevertheless, the possibility of an internal power source can not yet
be eliminated out of hand.  Indeed, such extra heating may emerge as 
another degree of freedom in the fits.  However, at present we find that Occam's Razor
and the arguments in \S\ref{metal_nocore} and \S\ref{core} obviate the necessity
for a central role for such an ad hoc core power of undetermined provenance. 

We end this section with a curious observation.  On 
Fig. \ref{fig:9}, we have placed gold points to indicate the approximate
core masses necessary to fit models having an extra internal
power source whose magnitude is an arbitrary, fixed, 
percentage (0.3\%) of L$_p$ (the same percentage for all the different L$_p$s).  These models
have solar-metallicity atmospheres, but no irradiation or $\Delta$R effects.
Even for these models, we see the same general trend of inferred 
core mass with stellar metallicity that was identified in \S\ref{core}.

\section{Ambiguity in Cooling from the Day/Night Sides}
\label{day}

The day/night difference in the cooling rates of strongly irradiated
planets remains the most uncertain aspect of all published theories.
If there is no modification of the heat flux at the radiative/convective
boundary on the night side due to heat redistribution at depth from the day side, 
and the planet cools on the night side as if isolated, then the nightside losses will
overwhelm the much smaller dayside losses and an extra heat source
(\S\ref{heat}) may well be required to explain those EGPs with the largest transit radii.  

In our default cooling model, we set the redistribution parameter, $f$,
defined in Burrows, Sudarsky, \& Hubbard (2003) and Burrows et al. (2004),
and used by other groups (e.g., Fortney et al. 2006; Chabrier et al. 2004), equal 
to 1/4.  This value signifies complete heat redistribution at depth and 
longitude-independent interior core fluxes outward.  The factor, $f$, 
influences the day/night temperature(T))/pressure(P) profile contrasts
only at high pressures near the radiative/convective boundary (at Rosseland $\tau$s
of $\sim$10$^6$; see Burrows et al. 2004).  At altitude, the day/night
contrast in an EGP's spectrum, formed at lower Rosseland $\tau$s of 0.1 to a few, can be large,
as suggested by the recent Ups And b light-curve data (Harrington
et al. 2006). However, at the same time, zonal winds at high optical
depths can still efficiently redistribute heat and entropy.
It is the T/P profile at depth that regulates core cooling.
Such efficient deep heat transport is suggested in the work of
Showman \& Guillot (2002) and Guillot \& Showman (2002), but is by
no means proven. Nevertheless, we make this assumption
in order to discover and explain the systematic features
across the family of known transiting EGPs.

As an aside, we note that many people think that rotation
is an efficient means to transport heat globally, and use Jupiter and Saturn
as examples. There is almost no latitude or longitude dependence
of the mid- and far-infrared emissions of either Jupiter or Saturn.
Despite the secant effect of the incident stellar flux, their emission temperatures
at these wavelengths are almost completely uniform.  However, it is not
rotation that smooths out these emissions, but the direct
heating of the convective regions of these planets by solar infrared
(Ingersoll 1976; Hubbard 1977; and Ingersoll and Porco 1978).
The radiative/convective boundary is at low optical depths
in these solar-system giants.  As a result, the solar heat directly absorbed
in the convective zone is efficiently redistributed throughout the planet's interior,
setting a uniform inner boundary for internal heat flux outward.
For closer-in EGPs, the radiative/convective boundary is
at greater depths and this mechanism does not operate. The upshot is
that for more strongly-irradiated EGPs, the mechanisms 
for longitudinal heat transport are more subtle, and problematic.
A number of groups are attempting to address this issue with multi-dimensional,
though approximate, numerical tools (Menou et al. 2003; Cho et al. 2003; 
Burkert et al. 2005; Cooper \& Showman 2005), but these efforts are only in their early stages.

\section{Discussion and Conclusions}
\label{conclusions}

In this paper, we have calculated the theoretical evolution of the 
radii of all fourteen of the known transiting giant planets
for a variety of assumptions concerning their atmospheric opacities,
inner core masses, and possible internal power sources.  We have
incorporated the effects of stellar irradiation and have customized 
such effects for each EGP and star.  Using measurements of their
ages, masses, and transit radii, we have sought to reconcile these
transit radii with theory.  While it can 
be difficult to fit each EGP definitively, looking at them collectively
can reveal important underlying features of the family as a whole.  
In doing so, we find that there are two, not one, radius anomalies.  Not only
are the radii of a subset of the known transiting EGPs larger
than expected from previous theory, but many of the other objects are smaller
than expected.  Unless all the atmospheres have only the default 
$\sim$solar-metallicity opacities, the $\Delta$R effect
can be ignored, and an internal power source whose magnitude is not correlated
in any obvious way with system parameters is operative, we conclude that the spread of 
measured radii is too large not to admit of a dual problem.   

We suggest that the larger EGPs can be explained
by invoking enhanced opacity atmospheres, that might 
be due only in part to enhanced metallicity, that naturally 
retain internal heat, and, hence, maintain their radii larger, longer. This can be done without
an extra internal power source, though such a source can not yet be
eliminated either as an important or a sub-dominant aspect of the 
theory for some irradiated EGPs.  We offer enhanced 
atmospheric opacities as a more straightforward explanation for the large-radius EGPs. 
Such an explanation, however, may require non-equilibrium chemistry 
and/or haze formation in the severe irradiation regimes in which transiting EGPs
find themselves and we have not provided in this paper a detailed 
chemical rationale for such altered atmospheres.

Furthermore, we suggest that the other anomaly, that of the small radii 
we find for the majority of the known transiting 
EGPs, can be explained simply by the presence of dense
cores and/or metal-rich envelopes in most, or all, of these fourteen objects.  
For no EGP orbiting a lower-metallicity star do we infer 
a large inner core.  Conversely, for no EGP orbiting the highest-metallicity
stars do we infer a small inner core. Moreover, the core masses
we find for EGPs transiting near-solar metallicity stars 
are close to those estimated for Jupiter and Saturn. 
Importantly, we derive a roughly montonically-increasing relationship 
between the stellar metallicity and the estimated core mass.
High stellar metallicity has been 
shown to correlate with the probability of the presence of an EGP
in the radial-velocity data (Fischer \& Valenti 2005).  In this paper, we find that
high stellar metallicity may also imply large inner cores and/or metal-rich
envelopes.  These twin correlations may speak to the mechanism of EGP formation
and are in keeping with the core-accretion model of their origin. 

\begin{figure}
\includegraphics[angle=0,width=15cm]{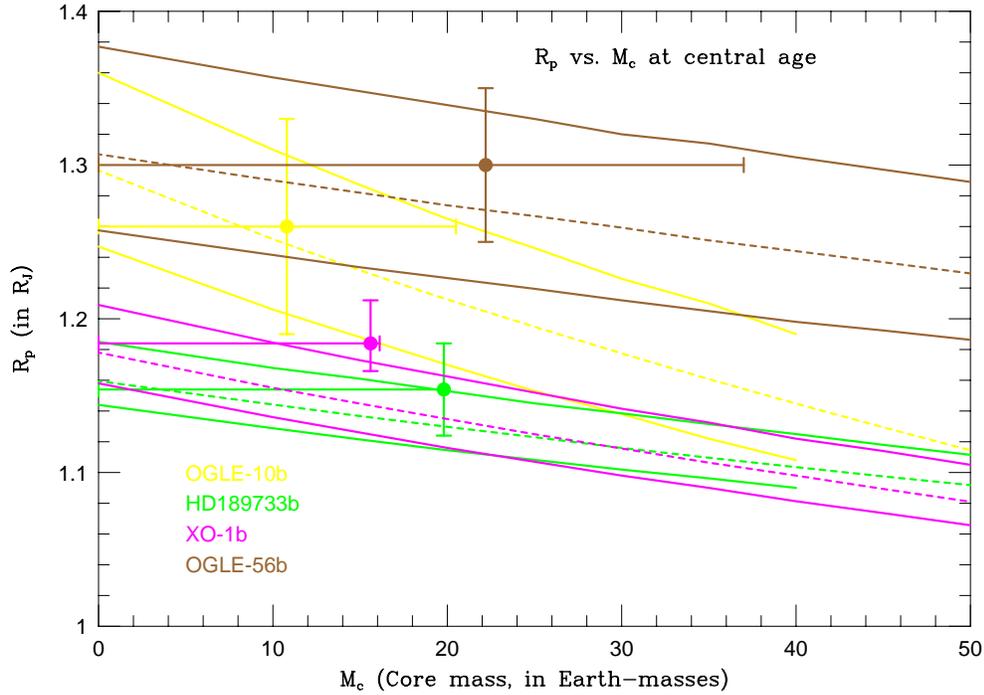}
\caption{Theoretical R$_p$ (in \rj) versus core mass, M$_c$ (in Earth masses), for
OGLE-TR-10b (yellow), OGLE-TR-56b (gold), HD189733b (green), and XO-1b (purple).
The lines are for solar, 3$\times$solar, and 10$\times$solar
atmosphere models and the 3$\times$solar models are dashed.  Central values
of the estimated stellar ages (Table 2) are assumed.  The measured radii of these 
transiting EGPs, along with 1-$\sigma$ error bars (vertical), are given.
The dots are put arbitrarily at core masses that represent 3$\times$solar metallicity
for the given EGP's measured mass and the rightmost extent of the horizontal ``error bars"
is placed at 3$\times${\it stellar} metallicity masses.  If 
the central value of the estimated stellar metallicity is below 
solar (as for HD189733b), the line is truncated at the dot.
Note that to construct the dots the heavy-element 
fractions of the atmosphere and of the envelope/core are here set equal. 
See text in \S\ref{core} for explanations and a discussion.
\label{fig:8}}
\end{figure}

\begin{figure}
\includegraphics[angle=0,width=9.0cm]{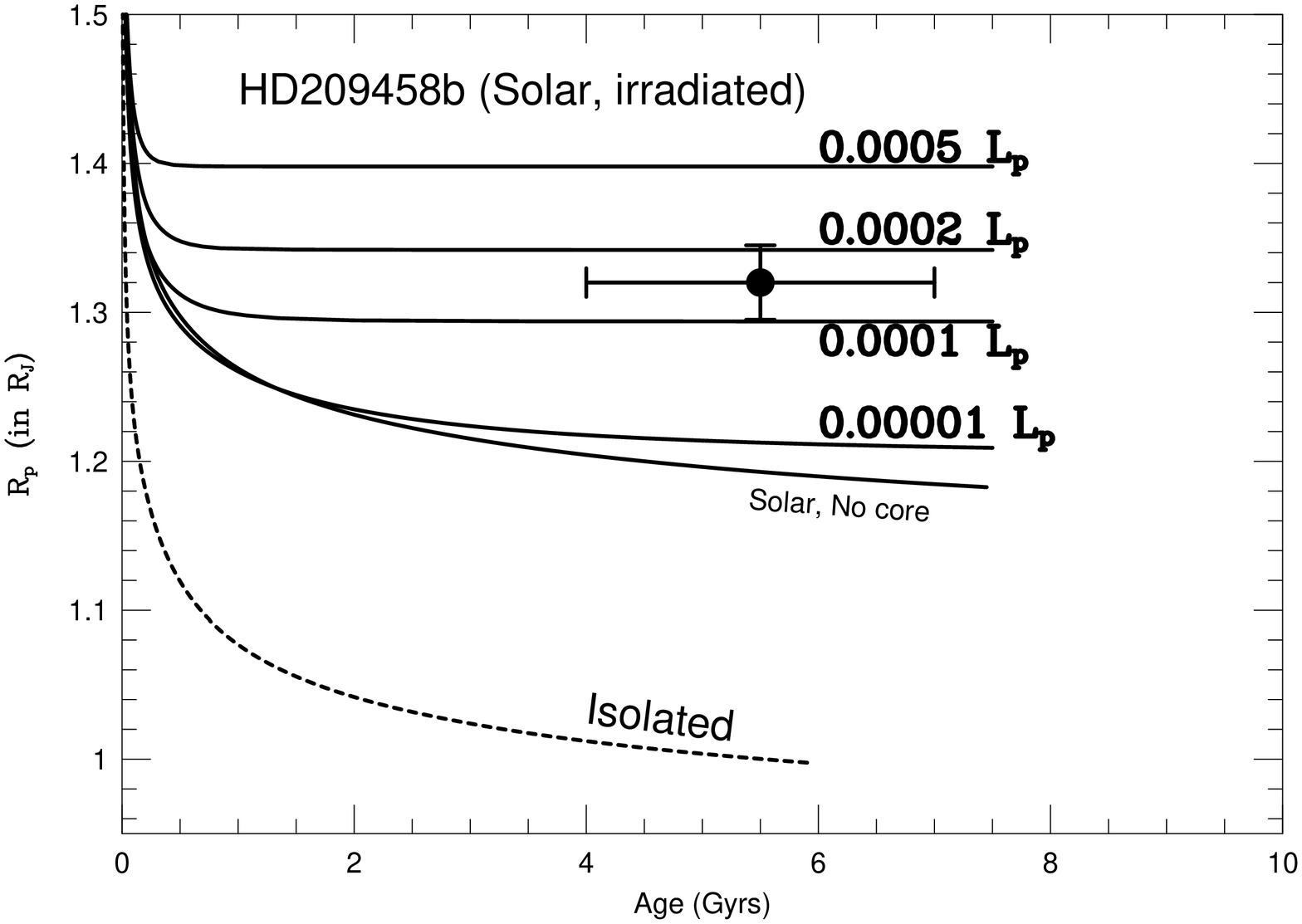}
\includegraphics[angle=0,width=9.0cm]{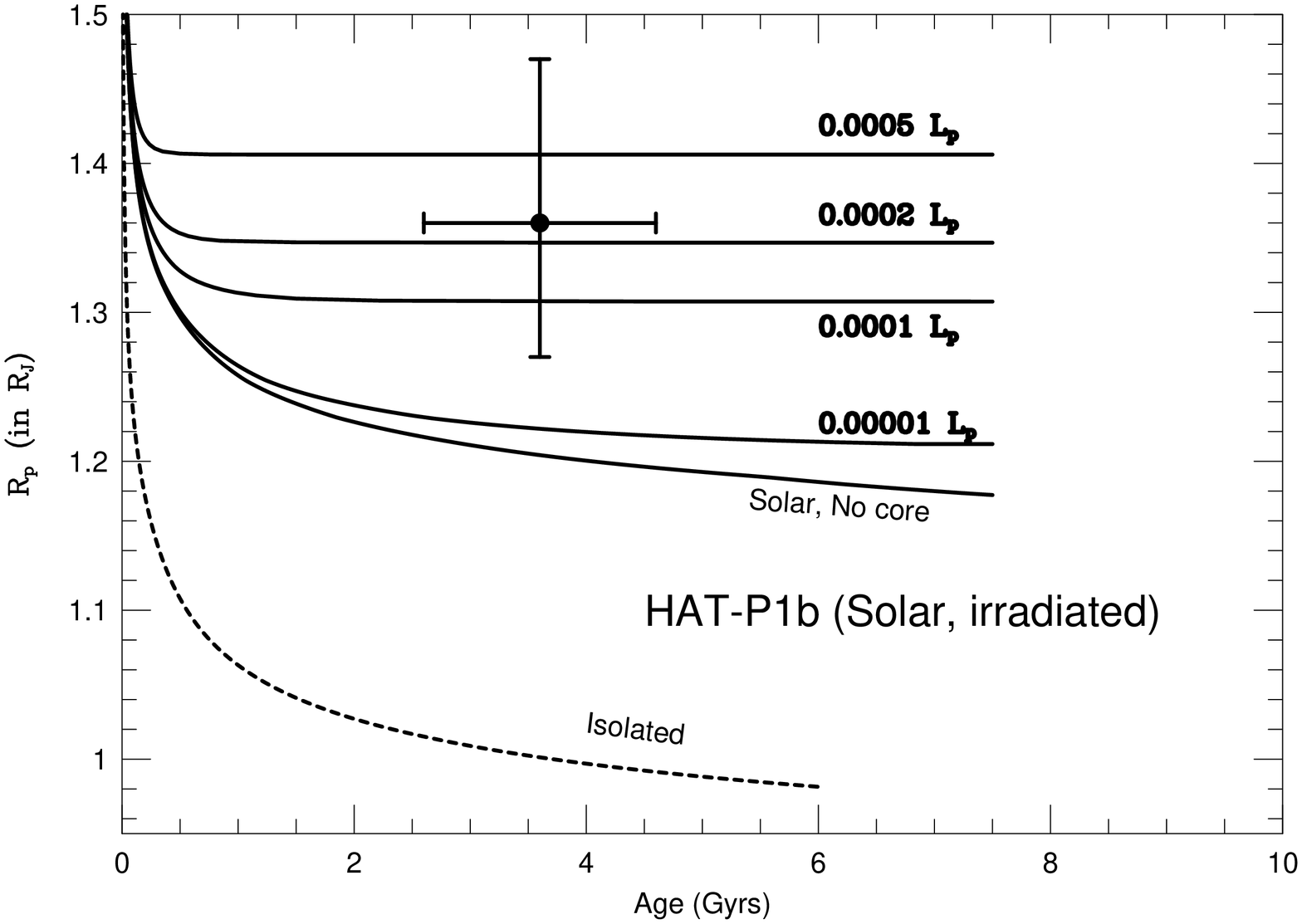}
\includegraphics[angle=0,width=9.0cm]{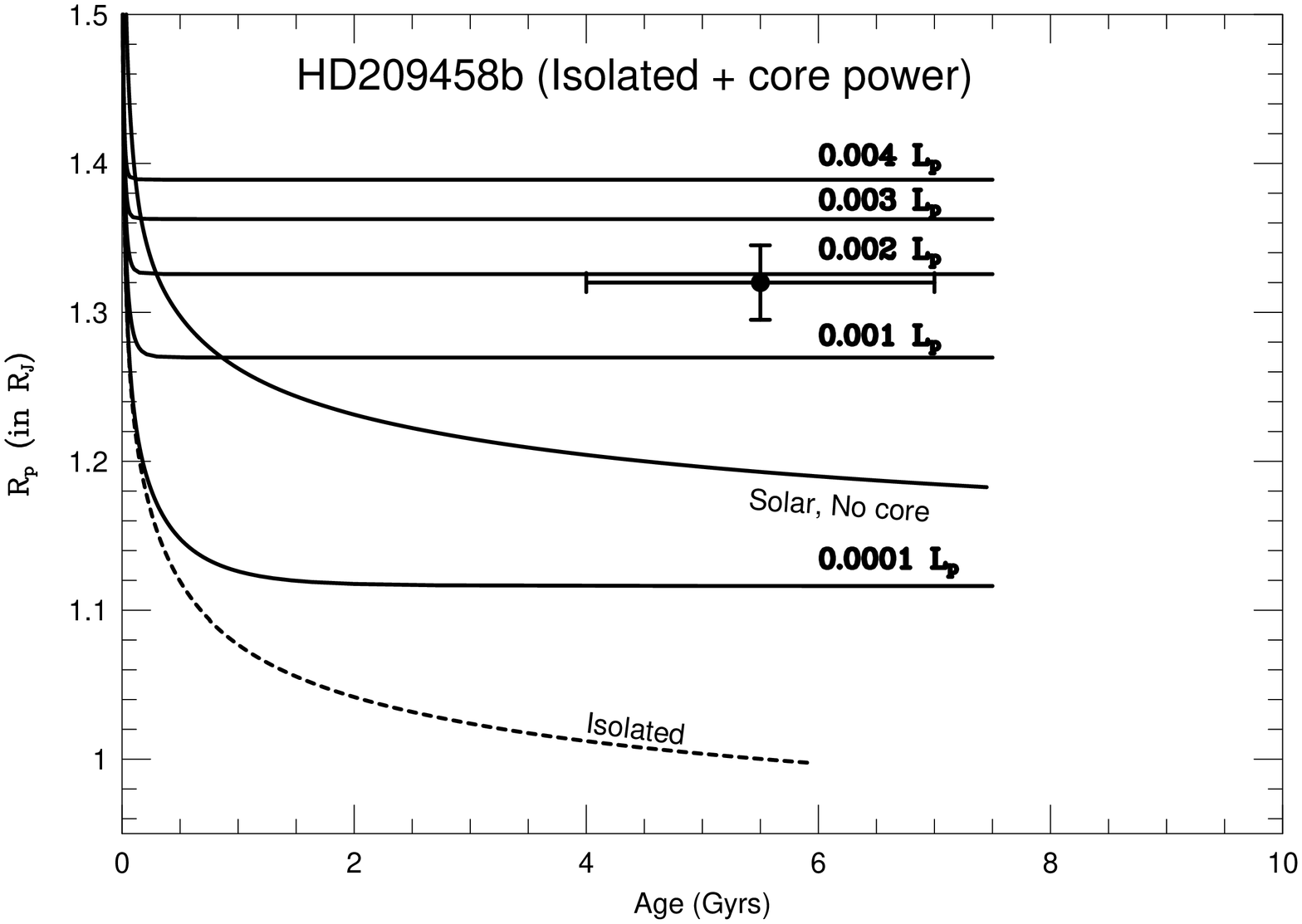} 
\includegraphics[angle=0,width=9.0cm]{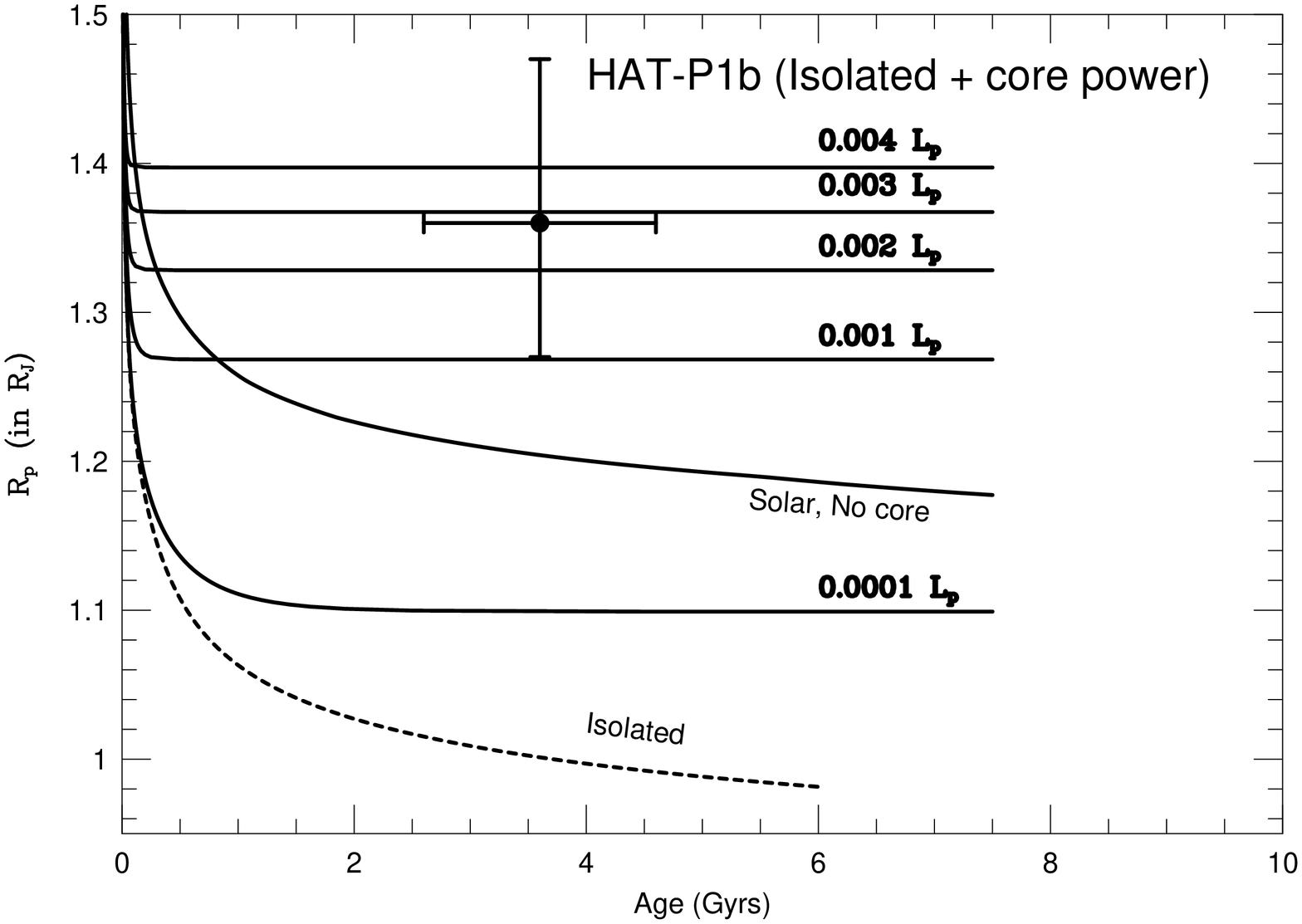}
\caption{{\bf Top:} Theoretical radii of HD209458b (left) and HAT-P1b (right), in units of \rj, versus
age (in Gyrs) for different values of an hypothesized core power for irradiated 
atmospheres with solar-metallicity opacities and no solid inner core.  The core power lines are identified
by the fraction of L$_p$, the total stellar power intercepted by the planet.
For HD209458b, this is $\sim$7.86$\times 10^{-5}$ L$_{\odot}$ and for HAT-P1b it is 
$\sim$5.29$\times 10^{-5}$ L$_{\odot}$ (Table 4).  The dashed lines are 
evolutionary trajectories for the respective isolated EGPs with solar-metallicity atmospheres
without irradiation and without the $\Delta$R term. The lines identified with the words ``Solar, No core"
are the no-core/irradiated/solar-atmosphere models of Fig. \ref{fig:8}.
{\bf Bottom:} Same as for the top, but for isolated atmospheres 
with solar opacities, without irradiation, and with the indicated 
core powers.  Note that the core powers required in this case are 
much larger than in the irradiated case depicted in the top two panels.
See \S\ref{heat} for a discussion.
\label{fig:10}}
\end{figure}

There are a number of caveats to our conclusions.  First is the uncertainty
concerning the nightside cooling.  If there is no means by which cooling
of the interior can be stanched by heat redistribution at depth from the dayside 
(Burrows et al. 2004), then an extra power source might be required for the larger radii.
Second is the wild card of rotation.  Since close-in EGPs are no doubt in synchronous
rotation at periods larger than those of Jupiter and Saturn, the effects
of rotation will result in no more than a few percent expansion, but have 
not yet been included in our analysis. Furthermore, centrifugal expansion is 
most manifest in the transit plane. Third is the possibility of delayed migration
of some of the planets.  If migration were to take many tens of millions of
years (Murray et al. 1998), then the planet might have had time to cool and shrink as if
in isolation, without the benefit of the effects of irradiation.  Subsequent irradiation
when in extremis could not reinflate the core (Burrows et al. 2000).  Fourth is
the fact that we have merely motivated altered chemistry in the atmospheres
of these severely irradiated EGPs, and have not demonstrated the required
chemistry, nor the opacity-enhancing effects.  High-metallicity atmospheres
in themselves would be adequate, but if these were accompanied by envelopes
with similar metallicities, the radius-increasing effect can be partially of wholly cancelled.
As Fig. \ref{fig:8} demonstrates, in many, though not all, 
of the cases the enhanced opacity effect of super-solar
metallicity in the atmosphere can still trump the shrinkage effect of the same
metallicity in the envelope. Super-solar metallicity in the atmosphere, expected generically
for EGPs, can still be part of the solution to the large-radius problem. 
However, in this study we have decoupled the two and future detailed work on
UV-driven chemistry, the opacities of strongly irradiated and synchronously-rotating
atmospheres,  and the equation of state for general mixtures is clearly needed.  Fifth
is the possibility that the heavy elements and the dominant absorbing compounds 
of the atmosphere might settle gravitationally, thereby depleting it of its
high-opacity components. Without these species, the high-opacity effect
that we suggest may be instrumental in explaining the largest EGP radii
would be compromised.  However, mixing due to the vigorous shear
motions caused by the zonal winds anticipated throughout these regions may
in fact be adequate to ensure an unstratified atmosphere. Nevertheless,
relevant calculations to estimate such mixing are warranted.  Finally and
sixth are the remaining ambiguities in system age, EGP radius, and stellar metallicity.
The inferred core masses, or range of core masses, and the fits to the
larger-radius EGPs depend upon those parameters.  Our results could
be more robust or less robust, depending upon the eventual values of these quantities.

\acknowledgments

We thank Maki Hattori, Kat Volk, Christopher Sharp,  
Mike Cushing, Dimitar Sasselov, and Drew Milsom for helpful discussions
and Francis O'Donovan for sharing preliminary data 
on TrES-2 in advance of publication. We also thank the 
referee for his careful reading of the manuscript.
This study was supported in part by NASA grants NNG04GL22G and NNG05GG05G and 
through the NASA Astrobiology Institute under Cooperative 
Agreement No. CAN-02-OSS-02 issued through the Office of Space
Science.

{}

\clearpage

\begin{table*}
\small
\begin{center}
\caption{Transiting Planet Data$^1$}
\tablewidth{17.0cm}
\begin{tabular}{ccccccc}
\hline\hline
    Planet&      a   & Period    &  $M_{p}$                & $R_{p}$   &F$_{p}$& Ref.   \\
               &   (AU)     &  (day)      &  ($M_{J}$)  & ($R_{J}$)   &(${\rm 10^9\ erg\ cm^{-2}\ s^{-1}}$)&    \\
\hline
   OGLE-TR-56b &  0.0225  & 1.2119 &  $1.29\pm0.12         $ & $1.30\pm0.05            $& 4.112 & 1,2,3,4,5 \\
  OGLE-TR-113b &  0.0229  & 1.4325 &  $1.32\pm0.19         $ & $1.09\pm0.03            $& 0.739 & 1,2,4,6,7,8 \\
  OGLE-TR-132b &  0.0306  & 1.6899 &  $1.19\pm0.13         $ & $1.13\pm0.08            $& 4.528 & 2,7,9 \\
       WASP-2b &  0.0307  & 2.1522 &  $0.88\pm0.11         $ & $1.04\pm0.06            $& 0.579 & 10,11  \\
    HD 189733b &  0.0313  & 2.2186 &  $1.15\pm0.04         $ & $1.15\pm0.03            $& 0.468 & 4,12,13 \\
       TrES-2  &  0.0367  & 2.4706 &  $1.28^{+0.09}_{-0.04}$ & $1.24^{+0.09}_{-0.06}   $& 1.150 & 14      \\
       WASP-1b &  0.0382  & 2.5199 &  $0.87\pm0.07         $ & $1.40\pm0.08            $& 2.488 & 10,15  \\
       TrES-1  &  0.0393  & 3.0301 &  $0.75\pm0.07         $ & $1.08\pm0.3             $& 0.428 & 1,2,4,16,17 \\
   OGLE-TR-10b & 0.0416   & 3.1013 &  $0.63\pm0.14         $ & $1.26\pm0.07            $& 1.344 & 1,2,4,5,18 \\
    HD 149026b &   0.042  & 2.8766 &  $0.36\pm0.03         $ & $0.73\pm0.03            $& 2.089 & 4,19    \\
    HD 209458b &   0.045  & 3.5247 &  $0.64\pm0.06         $ & $1.32\pm0.03            $& 1.074 & 4,20,21 \\
  OGLE-TR-111b &   0.047  & 4.0144 &  $0.52\pm0.13         $ & $1.07\pm0.05            $& 0.248 & 1,2,4,22 \\
         XO-1b &  0.0488  & 3.9415 &  $0.90\pm0.07         $ & $1.18^{+0.03}_{-0.02}   $& 0.485 & 23,24    \\
     HAT--P-1b &  0.0551  & 4.4653 &  $0.53\pm0.04         $ & $1.36^{+0.11}_{-0.09}   $& 0.681 & 25      \\
\hline
\end{tabular}
\tablenotetext{1}{Data, plus representative references, for the fourteen known transiting EGPs
with measured M$_p$ and R$_p$.  The list is in order of increasing semi-major axis.  F$_p$ is the 
stellar flux at the planet's substellar point, given the stellar luminosities provided in Table 2.
}
\bigskip
\tablerefs{
(1) Santos et al. (2006a),
(2) Santos et al. (2006b),
(3) Vaccaro \& Van Hamme (2005),
(4) Melo et al. (2006),
(5) Pont et al. (2006),
(6) Gillon et al. (2006),
(7) Bouchy et al. (2004),
(8) Konacki et al. (2004),
(9) Moutou et al. (2004),
(10) Cameron et al. (2006),
(11) Charbonneau et al. (2006c),
(12) Bouchy et al. (2005),
(13) Bakos et al. (2006a),
(14) O'Donovan et al. (2006),
(15) Shporer et al. (2006),
(16) Alonso et al. (2004),
(17) Winn, Holman, \& Roussanova  (2006),
(18) Holman et al. (2005),
(19) Sato, et al. (2005),
(20) Santos, Israelian, \& Mayor (2004),
(21) Knutson et al. (2006),
(22) Winn, Holman, \& Fuentes (2006),
(23) Holman et al. (2006),
(24) McCullough et al. (2006),
(25) Bakos et al. (2006b)
}
\label{t1}
\end{center}
\end{table*}


\begin{table*}
\small
\begin{center}
\caption{Data on Parent Stars$^1$}
\tablewidth{17.0cm}
\begin{tabular}{llllllllll}
\hline\hline
   Star & Sp.T.  & R$_*$                   &$T_{eff}$&$\log g$&[Fe/H]$_*$ & M    & L$_*$     & Age             & Dist  \\
              &        &  ($R_{\odot}$)             &  (K)   & (cgs)  &       &($M_{\odot}$)&($L_{\odot}$) &(Gyr)&(pc) \\
\hline
   OGLE-TR-56 &      G & $1.32\pm0.06            $& 6119 & 4.21 &  0.25 & 1.04 & 2.20   &$2.5^{+1.5}_{-1.0} $& 1600 \\
  OGLE-TR-113 &      K & $0.77\pm0.02            $& 4804 & 4.52 &  0.15 & 0.78 & 0.29  &$5.35\pm{4.65}    $&  550 \\
  OGLE-TR-132 &      F & $1.43\pm0.10            $& 6411 & 4.86 &  0.43 & 1.35 & 3.12   &$1.25\pm{0.75}    $& 2200 \\
       WASP-2 &    K1V & $0.81\pm0.03          $& 5200 & 4.50  &   $\cdots$ & 0.79 & 0.44 & $\cdots$ & $\cdots$  \\
    HD 189733 &  K1.5 & $0.76\pm0.02          $& 5050 & 4.53 & $-0.03$ & 0.82 & 0.34  &$5.25\pm{4.75}  $& 19.3 \\
       TrES-2 &    G0V & $1.00^{+0.06}_{-0.04}   $& 5960 & 4.40 & $-0.15$ &1.08& 1.14 &$7.2^{+1.8}_{-7.1}  $& $\cdots$ \\
       WASP-1 &    F7V & $1.42\pm0.07          $& 6200 & 4.30  &  $\cdots$   & 1.15 & 2.67   &  $\cdots$ &  $\cdots$   \\
       TrES-1 &    K0V & $0.81\pm0.02           $& 5226 & 4.40 &  0.06 & 0.88 & 0.49  &$4.0\pm{2.0}      $&  143 \\
   OGLE-TR-10 & G & $1.16\pm0.06            $& 6075 & 4.54 &  0.28 & 1.02 & 1.65   &$2.0\pm1.0        $& 1300 \\
    HD 149026 &  G0 IV & $1.45\pm0.10            $& 6147 & 4.26 &  0.36 & 1.3  & 2.71   &$2.0\pm{0.8}      $& 78.9 \\
    HD 209458 &   G0 V & $1.13\pm0.02            $& 6117 & 4.48 &  0.02 & 1.10& 1.60   &$5.5\pm{1.5}      $&   47 \\
  OGLE-TR-111 & G/K & $0.83\pm0.03          $& 5044 & 4.51 &  0.19 & 0.81 & 0.40  &$5.55\pm{4.45}    $& 1000 \\
         XO-1 &    G1V & $0.93^{+0.02}_{-0.01}$& 5750 & 4.53 & 0.015 & 1.00 & 0.85  &$4.6\pm{2.3}      $&  200 \\
     HAT-P-1 &    G0V & $1.15^{+0.10}_{-0.07}   $& 5975 & 4.45 &  0.13 & 1.12 & 1.52    &$3.6\pm{1.0}      $&  139 \\
\hline
\end{tabular}
\tablenotetext{1}{A compilation of the physical parameters derived for the parents of the known
transiting EGPs.  The error bars have been rounded from those found in the literature.  The ages, the
least well-known quantities, should be taken with caution, and those for WASP-1b and WASP-2b, since
unpublished, have been omitted. The stellar metallicities are given without error bars, which should be assumed large,
and are omitted for WASP-1b and WASP-2b for the same reason their ages are absent. Due to their great
distances (rightmost column), the stellar types of the OGLE objects are not well constrained.
Refer to Table 1 for the corresponding references.}
\label{t2}
\end{center}
\end{table*}

\clearpage

\begin{table*}
\small
\begin{center}
\caption{Approximate Inferred Core Mass Ranges given Central Age Estimates$^1$}
\tablewidth{20.0cm}
\begin{tabular}{ccccc}
\hline\hline
    Planet&  Solar   & 3$\times$Solar   &  10$\times$Solar & [Fe/H]$_*$  \\
\hline
   OGLE-TR-56b &  ($\cdots$) {\bf 0} (15)   & (0) {\bf 10} (25)     & (0) {\bf 20} (40)  & 0.25  \\
  OGLE-TR-113b &  (20) {\bf 60} (90)  & (40) {\bf 70} (115)     & (60) {\bf 80} (120)  & 0.15 \\
  OGLE-TR-132b &  (40) {\bf 85}($\cdots$)  & (75) {\bf 100} ($\cdots$)     & (90) {\bf 110} ($\cdots$) & 0.43 \\
       WASP-2b & $\cdots$   & $\cdots$     & $\cdots$  & $\cdots$ \\
    HD 189733b & ($\cdots$) {\bf 0} (15)   & (0) {\bf 5} (25)     & (0) {\bf 20} (40)  & $-$0.03 \\
       TrES-2  & ($\cdots$) {\bf 0} ($\cdots$)   & ($\cdots$) {\bf 0} ($\cdots$)    & (0) {\bf 15} ($\cdots$)  & $-$0.15 \\
       WASP-1b & $\cdots$   & $\cdots$     & $\cdots$  & $\cdots$ \\
       TrES-1  & (0) {\bf 35} (55)   & (10) {\bf 42} (65)     & (20) {\bf 55} (70)  & 0.06 \\
   OGLE-TR-10b & ($\cdots$) {\bf 0} (15)  & (0) {\bf 10} (25)     & (0) {\bf 20} (40)  & 0.28 \\
    HD 149026b & {\bf 80}   & {\bf 90}     & {\bf 110}  & 0.36 \\
    HD 209458b & ($\cdots$) $\cdots$ ($\cdots$)   & ($\cdots$) $\cdots$ ($\cdots$)     & ($\cdots$) $\cdots$ ($\cdots$)  & 0.02  \\
  OGLE-TR-111b & (10) {\bf 22} (37)   &  (13) {\bf 27} (42)    & (20) {\bf 35} (50)  & 0.19  \\
         XO-1b & ($\cdots$) {\bf 0} ($\cdots$)   & (0) {\bf 0} (10) & (0) {\bf 10} (20)  & 0.015 \\
     HAT--P-1b & ($\cdots$) $\cdots$ ($\cdots$)   & ($\cdots$) $\cdots$ ($\cdots$)     & ($\cdots$) $\cdots$ (5)  & 0.13 \\
\hline
\end{tabular}
\tablenotetext{1}{This table provides estimates of the core masses (in Earth masses), or core mass ranges,
suggested by our models from the best approximate fits to the measured transit radii. 
The ``best fits" for the measured radii are given in {\bf bold}, while the core masses for $+$1-$\sigma$
and $-$1-$\sigma$ radii are given in parentheses to the left and right, respectively.
When no value is given in parentheses, such a value would be meaningless.  For HD149026b, we
provide only the central model estimates.  Since there are no published values for the ages
of WASP-1b and WASP-2b, core mass estimates for them have not been provided. Central
values of the stellar metallicity estimates are provide in the last column 
(see Table 2).  As the table headings imply, such 
estimates depend upon the atmospheric opacities.
Since core mass and atmospheric opacity act on the transit radius in opposite senses,
the larger the opacity, the larger the core needed to compensate. Remaining
large uncertainties in the planet ages, particularly for young ages, and the significant 
error bars in the planet radii translate into weaker constraints on the core masses than
one would like. The upshot is uncertainty and more degrees-of-freedom for the theoretical fits.  Nevertheless, this table
provides the range and basic systematics in the current family of known transiting EGPs for the
cores needed to explain in broad outline the measured transit radii.  See the text in \S\ref{core} for a discussion
of the issues involved and some conclusions from this table. See also Fig. \ref{fig:9}.}  
\label{t3}
\end{center}
\end{table*}

\clearpage

\begin{table*}
\small
\begin{center}
\caption{Internal Power That Would Be Necessary$^1$}
\tablewidth{20.0cm}
\begin{tabular}{ccccc}
\hline\hline
    Planet&  Power (Iso)  & Power (Solar) &  L$_p$ & F$_p$ \\
              & (\% L$_p$) &  (\% L$_p$)  & (L$_{\odot}$) & ($10^9$\ ergs \ cm$^{-2}$\ s$^{-1}$)\\
\hline
   OGLE-TR-56b &  0.3  &   0.05   &  2.93$\times 10^{-4}$           & 4.112  \\
  OGLE-TR-113b &  0.02  &   $\cdots$       &  3.63$\times 10^{-5}$           & 0.739  \\
  OGLE-TR-132b &  0.01  &   $\cdots$       &  2.42$\times 10^{-4}$           & 4.528  \\
       WASP-2b &  0.005  &   $\cdots$       &  2.62$\times 10^{-5}$           & 0.579  \\
    HD 189733b &  0.13  &   $\cdots$     &  2.62$\times 10^{-5}$           & 0.468  \\
       TrES-2  &  0.4  &   0.03  &  7.42$\times 10^{-5}$           & 1.150  \\
       WASP-1b &  0.45  &  0.022  &  2.04$\times 10^{-4}$           & 2.488  \\
       TrES-1  &  0.025  &  $\cdots$      &  1.95$\times 10^{-5}$           & 0.428  \\
   OGLE-TR-10b &  0.075  &  $\cdots$      &  8.95$\times 10^{-5}$           & 1.344  \\
    HD 149026b &  $\cdots$   &            $\cdots$      &  4.61$\times 10^{-5}$           & 2.089  \\
    HD 209458b &  0.2  &   0.013  &  7.86$\times 10^{-5}$           & 1.074  \\
  OGLE-TR-111b &  0.03 &    $\cdots$      &  1.04$\times 10^{-5}$           & 0.248  \\
         XO-1b &  0.15  &   0.01  &  2.86$\times 10^{-5}$           & 0.485  \\
     HAT--P-1b &  0.3  &   0.025  &  5.29$\times 10^{-5}$           & 0.681  \\
\hline
\end{tabular}
\tablenotetext{1}{Some have suggested that the larger transit radii seen for some EGPs,
such as HD209458b, HAT-P1b, WASP-1b, might require an extra internal power source.
While not our preferred model (see \S\ref{heat} for a discussion), we provide in this table
the power (in percent of the intercepted stellar power, L$_p$, also given in this table 
for each EGP) that would be necessary to affect such inflation to the central 
measured value of the transit radius (Table 1) for two classes of models. As are 
the other tables, this table is in order of increasing orbital semi-major axis.  The first
class is for isolated, solar-metallicity, non-irradiated, EGPs (``Power (Iso)") 
and the second class is for our solar-metallicity irradiated models (``Power (Solar)").  
As can be seen, the latter class of models would require $\sim$ten 
times less extra internal power.  Also, many EGPs would ``require" no ($\cdots$) 
extra power, even for solar-metallicity atmospheres. 
Also provided is the stellar flux (F$_p$) at the substellar point of the planet (repeated from Table 1).
See text in \S\ref{heat} for a discussion.}
\label{t4}
\end{center}
\end{table*}

\clearpage

\clearpage

\clearpage

\clearpage

\clearpage

\clearpage

\clearpage

\clearpage

\end{document}